\documentclass[aps,prc,nofootinbib,superscriptaddress,eqsecnum,showpacs,preprint]{revtex4-1}
\pdfoutput=1

\usepackage{color}
\usepackage{graphicx}
\usepackage{bm} 
\usepackage{epsfig}
\usepackage[latin1]{inputenc}
\usepackage{float,amsmath}
\usepackage{amssymb}
\usepackage{wrapfig}

\newcommand{\eiso}{{\cal E}_{\rm iso}}
\newcommand{\piso}{{\cal P}_{\rm iso}}
\newcommand{\ahydro}{{\sc aHydro }}
\newcommand{\eavg}[1]{<\!#1\!>}

\begin{document}


\title{Boost-Invariant (2+1)-dimensional Anisotropic Hydrodynamics}


\author{Mauricio Martinez}
\affiliation{Departamento de F\'isica de Part\'iculas, Universidade de Santiago de Compostela,
E-15782 Santiago de Compostela, Galicia, Spain}

\author{Radoslaw Ryblewski} 
\affiliation{The H. Niewodnicza\'nski Institute of Nuclear Physics, Polish Academy of Sciences, PL-31342 Krak\'ow, Poland}

\author{Michael Strickland}
\affiliation{
  Physics Department, Gettysburg College\\
  Gettysburg, PA 17325 United States
}
\affiliation{
Frankfurt Institute for Advanced Studies\\
Ruth-Moufang-Strasse 1\\
D-60438, Frankfurt am Main, Germany
}


\begin{abstract}
We present results of the application of the anisotropic hydrodynamics ({\sc aHydro}) framework to (2+1)-dimensional 
boost invariant systems.  The necessary {\sc aHydro} dynamical equations are derived by taking moments of the 
Boltzmann equation using a momentum-space anisotropic one-particle distribution function.  We present a derivation of
the necessary equations and then proceed to numerical solutions of the resulting partial differential equations using both 
realistic smooth Glauber initial conditions and fluctuating Monte-Carlo Glauber initial conditions.  For this purpose we have 
developed two numerical implementations:  one which is based on straightforward integration of the resulting partial 
differential equations supplemented by a two-dimensional weighted Lax-Friedrichs smoothing in the case of fluctuating
initial conditions; and another that is based on the application of the Kurganov-Tadmor central scheme.  For our final 
results we compute the collective flow of the matter via the lab-frame energy-momentum tensor eccentricity as 
a function of the assumed shear viscosity to entropy ratio, proper time, and impact parameter.
\end{abstract}



\preprint{INT-PUB-12-014}

\maketitle 


\section{Introduction}

The goal of ultrarelativistic heavy ion collision experiments at the Relativistic Heavy Ion Collider 
at Brookhaven National Laboratory (RHIC) and the Large Hadron Collider (LHC) at CERN is to create a 
tiny volume of matter ($\sim$ 1000 fm$^3$) which has been heated to a temperature exceeding that 
necessary to create a quark-gluon plasma.  Early on it was shown that ideal relativistic hydrodynamics is
able to reproduce the soft collective flow of the matter and single particle spectra produced at RHIC 
\cite{Huovinen:2001cy,Hirano:2002ds,Tannenbaum:2006ch, Kolb:2003dz}.  
Based on this there was a concerted effort to develop a more systematic framework for describing the soft 
collective motion.  This effort resulted in a number of works dedicated to 
the development and application of relativistic viscous hydrodynamics 
to relativistic heavy ion collisions
\cite{Muronga:2001zk,Muronga:2003ta,Muronga:2004sf,Baier:2006um,Romatschke:2007mq,Baier:2007ix,%
Dusling:2007gi,Luzum:2008cw,Song:2008hj,%
Heinz:2009xj,El:2009vj,PeraltaRamos:2009kg,PeraltaRamos:2010je,Denicol:2010tr,Denicol:2010xn,%
Schenke:2010rr,Schenke:2011tv,Bozek:2011wa,Niemi:2011ix,Niemi:2012ry,Bozek:2012qs,Denicol:2012cn}.  

One of the weakness of the traditional viscous hydrodynamics approach 
is that it relies on an implicit assumption that the system 
is close to thermal equilibrium which implies that the system is also very close to being isotropic in momentum space. 
However, one finds during the application of these methods to relativistic heavy ion collisions that this 
assumption breaks down at the earliest times after the initial impact of the two nuclei due to large 
momentum-space anisotropies in the $p_T$-$p_L$ plane which can
persist for many fm/c \cite{Martinez:2009mf}.  In addition, one finds that near the transverse and longitudinal edges of 
the system these momentum-space 
anisotropies are large at all times \cite{Martinez:2009mf,Martinez:2010sd,Ryblewski:2010bs,%
Strickland:2011aa,Strickland:2011mw}.  Similar conclusions have been obtained in the context of strongly
coupled systems where it has been shown using the AdS/CFT correspondence one achieves viscous 
hydrodynamical behavior at times when the system
still possesses large momentum-space anisotropies and that these anisotropies remain large throughout the evolution 
\cite{Chesler:2008hg,Chesler:2009cy,Heller:2011ju,Heller:2012je,Heller:2012km,Wu:2011yd,Chesler:2011ds}.  
Based on these results one is motivated to obtain
a dynamical framework that can accommodate potentially large momentum-space anisotropies.

In this paper we follow up recent work which aims to extend the applicability of space-time evolution models
for the bulk dynamics of a quark-gluon plasma to situations in which there can be large momentum-space anisotropies.  
Initial studies along this direction focused on boost-invariant expansion in systems which
were transversally homogeneous \cite{Florkowski:2010cf,Martinez:2010sc}.  The motivation and conceptual
setup of Refs.~\cite{Florkowski:2010cf,Martinez:2010sc} were similar in the sense that they both relaxed the 
assumption of the system being nearly isotropic in momentum space; however, there was a key conceptual difference in 
the derivation of the resulting dynamical equations.  In
Ref.~\cite{Florkowski:2010cf} an entropy source was postulated which satisfied the minimal constraints
necessary in the limit of small momentum-space anisotropy and then the authors assumed a constant rate 
of isotropization regardless of the local typical momentum of the plasma constituents.  In 
Ref.~\cite{Martinez:2010sc} the equations of motion were derived by taking moments of the Boltzmann equation 
and supplemented by a requirement that in the limit of small momentum-space anisotropy these equations 
reproduced those of 2nd-order Israel-Stewart viscous hydrodynamics \cite{GLW,Israel:1979wp,Muronga:2006zx}.  
The result of this matching was that the relaxation rate of the system was necessarily proportional to the 
local hard momentum scale.\footnote{In this context the hard momentum scale corresponds to the typical average momentum 
scale of the particles of the system. When one has local isotropic thermal equilibrium, the average momentum scale corresponds to 
the temperature of the system.} This allowed the authors of Ref.~\cite{Martinez:2010sc} to smoothly match onto 
2nd-order viscous hydrodynamics when the system was nearly isotropic in momentum space.

The phenomenological consequence of these two different results for the relaxation rate is quite important.  If the 
relaxation rate is proportional to the local hard momentum scale, then one expects a slower relaxation to isotropy 
when the local hard momentum scale is reduced.  This occurs at late times in the one-dimensional case since the local 
hard momentum scale is dynamically lowered due to expansion.  Even more importantly, having a relaxation rate which
is proportional to the hard momentum scale has important consequences for the evolution of the matter near the longitudinal and 
transverse edges of the system where the local temperature is also initially lower.  The first demonstration
of this effect was in Ref.~\cite{Martinez:2010sd} which studied the one dimensional non-boost invariant evolution of a system
which was transversally homogeneous.  This work followed similar developments in Ref.~\cite{Ryblewski:2010bs} 
where a constant relaxation rate was assumed.  A comparison of the results of these two papers shows 
that one sees much larger momentum-space anisotropies at large spatial rapidity being developed if one 
uses a relaxation rate which is proportional to the local hard momentum scale.

Since these works were published, the anisotropic hydrodynamics methodology has been extended to 
include boost-invariant transverse dynamics \cite{Ryblewski:2011aq,Florkowski:2011jg}; however, these
papers once again assumed a fixed rate of relaxation to isotropy.  In this paper we study the effect of
using a more realistic relaxation rate which is proportional to the hard momentum scale \cite{Martinez:2010sc},
thereby allowing a smooth matching to 2nd-order viscous hydrodynamics.  We present a derivation of the necessary 
equations and then proceed to numerical solutions of the resulting partial differential equations using both realistic 
smooth Glauber initial conditions and fluctuating Monte-Carlo Glauber initial conditions.  For this purpose we have 
developed two numerical implementations:  one which is based on straightforward integration of the resulting partial 
differential equations supplemented by a two-dimensional weighted Lax-Friedrichs smoothing in the case of fluctuating 
initial conditions; and another that is based on the application of the Kurganov-Tadmor central scheme.  For our final 
results we compute the collective flow of the matter via the lab-frame energy-momentum tensor eccentricity as 
a function of the assumed shear viscosity to entropy ratio, proper time, 
and impact parameter.  We also present results for the dependence
of the momentum-space anisotropy in the full transverse plane and show that in regions where the temperature is
low one can develop sizable momentum-space anisotropies.  As a control test we compare with 2nd-order viscous
hydrodynamics in the limit of small shear viscosities and demonstrate that the \ahydro framework is able to reproduce
the temperature and flow profiles obtained from 2nd-order viscous hydrodynamics in this limit.  

The structure of the paper is as follows:  In Sec.~\ref{sec:kineticth} we introduce the tensor basis we will use
in the case that the system is anisotropic in momentum space and derive the partial differential equations
necessary for the dynamical evolution by taking moments of the Boltzmann equation.  In Sec.~\ref{sec:initialconditions} 
we present the types of smooth initial conditions we will use.  In
Sec.~\ref{sec:nummeth} we introduce the three numerical algorithms (centered differences, weighted LAX, and hybrid 
Kurganov-Tadmor) we will we use to solve the resulting partial differential equations.  In Sec.~\ref{sec:results} we
compare with 2nd-order viscous hydrodynamics for non-central collisions and present our final results.  In 
Sec.~\ref{sec:conclusions} we present our conclusions and a future outlook.  Finally, in three appendices we include
a comparison of entropy production in 2nd-order viscous hydrodynamics and {\sc aHydro}, some numerical checks of 
convergence etc., and a brief rederivation of the 0+1d Bjorken model using our tensor formalism.

\section{Kinetic theory approach to anisotropic hydrodynamics}
\label{sec:kineticth}

In this section we describe our theoretical framework for describing relativistic plasmas which are anisotropic in 
momentum-space. Our setup is based on the kinetic theory approach to non-equilibrium systems \cite{GLW}. There are 
different methods for constructing approximate solutions of the relativistic Boltzmann equation \cite{GLW}. The most 
well-known approach is due to Israel and Stewart \cite{Israel:1976tn,Israel:1979wp}.  In this approach one expands the 
distribution function around a local thermal equilibrated distribution function, $f_{\rm eq}(x,p)$,
in terms of a series of irreducible Lorentz tensors \footnote{We point out that in the 
original approach by Israel and Stewart, the decomposition basis is not orthogonal and therefore, the exact form of the 
transport coefficients cannot be obtained once the expansion is truncated. Recently, Denicol et al.~showed how to correct 
this and expand properly the distribution function in terms of a complete and orthogonal set of irreducible tensors 
of a particle with momentum $p^\mu$ \cite{Denicol:2012cn}.} of particle momentum $p^\mu$
\begin{eqnarray}
\label{canonical}
f(x,p) &=& \,f_{\rm eq}(x,p)\,(1+\phi(x,t) )\,, \nonumber\\
&=& f_{\rm eq}(x,p)\,(1+c(x,t) + c_\mu p^{\langle\mu\rangle} + c_{\mu\nu} p^{\langle\mu}p^{\nu\rangle}
	+ c_{\mu\nu\lambda}p^{\langle\mu}p^\nu p^{\lambda\rangle} + \ldots \; ) \, ,
\end{eqnarray}
where the angle brackets above stand for symmetrized tensors which are orthogonal to the
fluid four-velocity $u^\mu$ \cite{GLW,Denicol:2012cn}.
The thermal equilibrium distribution function has the functional form
\begin{equation}
f_{\rm eq}= \left[ \exp\!\left(\frac{p^\mu u_\mu (x)-\mu(x)}{T(x)}\right) + a\right]^{-1} \, ,
\end{equation}
where $a= \pm 1$ gives Fermi-Dirac or Bose-Einstein statistics and $a=0$ gives Maxwell-Boltzmann statistics. 

The distribution function (\ref{canonical}) is usually expanded until second order, i.e. just keeping the terms 1, 
$p^{\langle\mu\rangle}$, and $p^{\langle\mu}p^{\nu\rangle}$. An important aspect in the construction of irreducible 
tensor basis is the decomposition of the four-momentum $p^\mu$ of a particle in Minkowski space. One assumes the existence of a 
time-like normalized vector field $u^\mu(x)$ (which is identified with the fluid velocity) and an operator $\Delta_{\mu\nu}$ 
which is symmetric, traceless and orthogonal to $u^\mu(x)$ such that $p^\mu = E u^\mu +\Delta^{\mu\nu}p_\nu$ 
\cite{GLW,Denicol:2012cn}. This decomposition allows one to have an irreducible $n$th-rank  tensor basis  which is complete and 
orthogonal \cite{GLW,Denicol:2012cn}. 

An alternative but equivalent treatment for expanding the distribution function in terms of an irreducible $n$th-rank
tensor basis was developed by Anderson \cite{anderson:1116}. This method instead decomposes the 
four-momentum $p^\mu$ of a particle as  
\begin{equation}
\label{adecomp}
p^\mu = E u^\mu + \sum_{i=1}^{3} p_i x^\mu_i \, ,
\end{equation}
where $u^\mu$ is the fluid velocity and $x^\mu_i$ is a set of orthonormal vectors which are spacelike and orthogonal 
to $u^\mu$. With this decomposition one can also find a suitable irreducible tensor representation \cite{anderson:1116}. We will 
follow this decomposition closely since it is the most convenient  vector basis for a system which is anisotropic in 
momentum-space along some preferred direction(s). 

In the rest of this section, we use the vector basis decomposition (\ref{adecomp}) to construct 2nd-rank tensors. As a particular case, we construct the energy-momentum tensor for a (2+1)-dimensional boost invariant anisotropic plasma and derive the dynamical equations of motion by taking moments of the Boltzmann equation.  Our discussion is restricted to the case of vanishing chemical potential.

\subsection{Vector Basis}

In this paper we will concentrate on systems which possess a preferred direction associated with a single direction 
in momentum-space. It is possible to construct a tensor basis which allows for multiple anisotropy directions; however, 
we restrict our considerations to this simpler case since taking into account the momentum-space anisotropy along the 
beamline direction is of particular importance for heavy-ion phenomenology.  To begin, we will specify a tensor basis
which is completely general and not subject to any symmetry constraints and then add the necessary symmetry
constraints when needed.

A general tensor basis can be constructed by introducing four 4-vectors which in the local rest frame (LRF) are
\begin{eqnarray}
&&X^\mu_{0,{\rm LRF}} \equiv u^\mu_{\rm LRF} = (1,0,0,0) \nonumber \\
&&X^\mu_{1,{\rm LRF}} \equiv x^\mu_{\rm LRF} = (0,1,0,0) \nonumber \\
&&X^\mu_{2,{\rm LRF}} \equiv y^\mu_{\rm LRF} = (0,0,1,0) \nonumber \\
&&X^\mu_{3,{\rm LRF}} \equiv z^\mu_{\rm LRF} = (0,0,0,1) \, .
\label{eq:rfbasis}
\end{eqnarray}
These 4-vectors are 
orthonormal in all frames.  The vector $X^\mu_0$ is associated with the four-velocity of the local rest frame and
is conventionally called $u^\mu$ and one can also identify $X^\mu_1 = x^\mu$, 
$X^\mu_2 = y^\mu$, and $X^\mu_3 = z^\mu$ as indicated above.  We will use the two different 
labels for these vectors interchangeably depending on convenience since the notation with numerical 
indices allows for more compact expressions in many cases.  Note that, in the lab frame the three 
spacelike vectors $X^\mu_i$ can be written entirely in terms of $X^\mu_0=u^\mu$.  This is 
because $X^\mu_i$ can be obtained by a sequence of Lorentz transformations/rotations applied 
to the local rest frame expressions specified above.  We will return to this issue and construct
explicit lab-frame representations of these four-vectors later.

Finally, we point out that one can express the metric tensor itself in terms of these 4-vectors as
\begin{equation}
g^{\mu \nu}= X^\mu_0 X^\nu_0 - \sum_{i=1}^3 X^\mu_i X^\nu_i \, .
\label{eq:gbasis}
\end{equation}
In addition, the standard transverse projection operator which is 
orthogonal to $X^\mu_0$ can be rewritten in terms of the vector basis (\ref{eq:rfbasis}) as
\begin{equation}
\Delta^{\mu \nu} = g^{\mu\nu} - X^\mu_0 X^\nu_0 = - \sum_{i=1}^3 X^\mu_i X^\nu_i \, ,
\label{eq:transproj}
\end{equation}
such that $u_\mu \Delta^{\mu \nu} = u_\nu \Delta^{\mu \nu} = 0$.  We note that the spacelike
components of the tensor basis are eigenfunctions of this operator, i.e. $X_{i\mu} \Delta^{\mu \nu} = X^\nu_{i}$.

 
\subsection{2nd-rank Tensors}
A general rank two tensor can be decomposed using the 4-vectors $X_\alpha^\mu$.  In general there are sixteen possible terms
\begin{eqnarray}
A^{\mu\nu}(t,{\bf x}) &=& \sum_{\alpha,\beta=0}^3 c_{\alpha\beta} X^\mu_\alpha X^\nu_\beta \, , \nonumber\\
&=& c_{00}X^\mu_0X^\nu_0 + \sum_{i=1}^3 c_{ii} X^\mu_i X^\nu_i 
+ \sum_{\alpha,\beta=0 \atop \alpha\neq\beta}^3 c_{\alpha\beta} X^\mu_\alpha X^\nu_\beta \, , \nonumber \\
&=& c_{00} g^{\mu \nu}  + \sum_{i=1}^3 \underbrace{(c_{ii}+c_{00})}_{\equiv \, d_{ii}} X^\mu_i X^\nu_i 
+ \sum_{\alpha,\beta=0 \atop \alpha\neq\beta}^3 c_{\alpha\beta} X^\mu_\alpha X^\nu_\beta \, ,
\end{eqnarray}
where it is understood that the coefficients $c_{\alpha\beta}$ now contain all of the space-time dependence.


\subsection{2nd-rank symmetric Tensors}

If a two tensor is symmetric under the interchange of $\mu$ and $\nu$ then $c_{\alpha\beta} = c_{\beta\alpha}$ and we can write
\begin{equation}
A^{\mu\nu}(t,{\bf x}) = c_{00} g^{\mu \nu}  + \sum_{i=1}^3 d_{ii} X^\mu_i X^\nu_i 
+ \sum_{\alpha,\beta=0 \atop \alpha>\beta}^3 c_{\alpha\beta} (X^\mu_\alpha X^\nu_\beta+X^\mu_\beta X^\nu_\alpha) \, .
\label{eq:symtensor}
\end{equation}
and there are only then ten independent terms.


\subsubsection{Energy-Momentum Tensor for Ideal Hydrodynamics}

Since the energy-momentum tensor is a symmetric tensor of 2nd-rank, Eq.~(\ref{eq:symtensor}) can be used 
\begin{equation}
T^{\mu\nu}(t,{\bf x}) = t_{00} g^{\mu \nu}  + \sum_{i=1}^3 t_{ii} X^\mu_i X^\nu_i 
+ \sum_{\alpha,\beta=0 \atop \alpha>\beta}^3 t_{\alpha\beta} (X^\mu_\alpha X^\nu_\beta+X^\mu_\beta X^\nu_\alpha) \, ,
\label{eq:emtensorgen}
\end{equation}
where we have relabeled the coefficients for this purpose.
In the local rest frame we can identify the basis vectors via (\ref{eq:rfbasis}) and we have that $T^{00}_{\rm LRF} = {\cal E}$ and 
$T^{ii}_{\rm LRF} = {\cal P}_i$ where ${\cal E}$ is the energy density and ${\cal P}_i$ is the pressure in $i$-direction and all other
components vanish.  
If the system is locally isotropic as is the case for ideal hydrodynamics then ${\cal P}_i \equiv {\cal P}$.  From (\ref{eq:emtensorgen}) 
we have $T^{00}_{\rm LRF} = {\cal E} = t_{00}$ and $T^{ii}_{\rm LRF} = {\cal P} = -t_{00} + t_{ii}$ and since all off-diagonal
components vanish we have $t_{\alpha\beta}$ = 0 for all $\alpha \neq \beta$.  This allows us to write
\begin{eqnarray}
T^{\mu\nu}(t,{\bf x}) &=& {\cal E} g^{\mu \nu}  + ({\cal P} + {\cal E}) \sum_{i=1}^3 X^\mu_i X^\nu_i \, , \nonumber \\
&=& {\cal E} g^{\mu \nu}  + ({\cal P} + {\cal E}) (X^\mu_0 X^\nu_0 - g^{\mu \nu}) \, \nonumber  \\
&=& ({\cal E} + {\cal P}) X^\mu_0 X^\nu_0  - {\cal P} g^{\mu \nu} \, ,
\label{eq:emtensorisotropic}
\end{eqnarray}
where in going from the first to second line we have used Eq.~(\ref{eq:gbasis}).
Using the conventional notation that $X^\mu_0 = u^\mu$ we obtain
\begin{equation}
T^{\mu\nu} = ({\cal E} +{\cal P}) u^\mu u^\nu  - {\cal P} g^{\mu \nu} \, ,
\label{eq:idealtmunu}
\end{equation}
in agreement with the expected result.  For later use we also note that
\begin{equation}
{T^\mu}_\mu \equiv {\cal T} = {\cal E} - 3 {\cal P} \, .
\label{eq:idealanomaly}
\end{equation}
%


\subsubsection{Energy-Momentum Tensor for Azimuthally-Symmetric Anisotropic Hydrodynamics}

In the bulk of this paper we will consider systems for which the momentum-space particle distribution
is azimuthally symmetric while the rotational symmetry in the $p_\perp$-$p_L$ plane is broken.  From
here on we will refer to this as  ``azimuthally-symmetric'' which only implies an assumed symmetry in 
momentum-space and not in configuration space.  In the
case of azimuthally-symmetric anisotropic hydrodynamics we have
\begin{eqnarray}
T^{00}_{\rm LRF} &=& {\cal E} = t_{00} \nonumber \, , \\
T^{xx}_{\rm LRF} &=& {\cal P}_\perp = -t_{00} + t_{11}\nonumber \, ,  \\
T^{yy}_{\rm LRF} &=& {\cal P}_\perp = -t_{00} + t_{22}\nonumber \, ,  \\
T^{zz}_{\rm LRF} &=&  {\cal P}_L = -t_{00} + t_{33} \, ,
\end{eqnarray}
and due to the azimuthal symmetry in momentum-space we must have $ t_{11}= t_{22}$ which gives 
four equations for our four unknowns.  Solving for the coefficients $t$ one obtains
\begin{eqnarray}
T^{\mu\nu}(t,{\bf x}) &=& {\cal E} g^{\mu \nu}  + ({\cal P}_\perp + {\cal E}) \sum_{i=1}^2 X^\mu_i X^\nu_i 
	+ ({\cal P}_L + {\cal E}) X^\mu_3 X^\nu_3 \, , \nonumber \\
&=& {\cal E} g^{\mu \nu}  + ({\cal P}_\perp + {\cal E}) \sum_{i=1}^3 X^\mu_i X^\nu_i 
	+ ({\cal P}_L - P_\perp) X^\mu_3 X^\nu_3 \, , \nonumber \\	
&=& {\cal E} g^{\mu \nu}  + ({\cal P}_\perp + {\cal E}) (X^\mu_0 X^\nu_0 - g^{\mu \nu}) 
	+ ({\cal P}_L - P_\perp) X^\mu_3 X^\nu_3 \, \nonumber  \\
&=& ({\cal E} + {\cal P}_\perp) X^\mu_0 X^\nu_0  - {\cal P}_\perp g^{\mu \nu} 
	+ ({\cal P}_L - P_\perp) X^\mu_3 X^\nu_3\, .
\label{eq:emtensorspheroidal}
\end{eqnarray}
Relabeling $X^\mu_0 = u^\mu$ and $X^\mu_3 = z^\mu$ to agree more closely 
with the notation of Ref.~\cite{Florkowski:2011jg} we obtain
\begin{equation}
T^{\mu\nu} = ({\cal E} +{\cal P}_\perp) u^\mu u^\nu  - {\cal P}_\perp g^{\mu \nu}+ ({\cal P}_L - {\cal P}_\perp) z^\mu z^\nu \, ,
\label{eq:speroidaltmunu}
\end{equation}
which in the limit that ${\cal P}_\perp = {\cal P}_L \equiv {\cal P}$ reduces to (\ref{eq:idealtmunu}).
We again note for later use that
\begin{equation}
{T^\mu}_\mu \equiv {\cal T} = {\cal E} - 2 {\cal P}_\perp  - {\cal P}_L\, .
\label{eq:spheroidalanomaly}
\end{equation}


\subsection{Explicit Forms of the Basis Vectors}

In the lab frame the three spacelike vectors $X^\mu_i$ can be written entirely 
in terms of $X^\mu_0=u^\mu$.  This is because $X^\mu_i$ can be obtained by a sequence 
of Lorentz transformations/rotations applied to the local rest frame expressions specified above. 
To go from the lab frame to LRF we can apply a boost along the $z$-axis followed by a rotation 
around the $z$-axis and finally a boost along the $x$-axis, i.e. $u_{\rm LRF} = L_x(\psi) 
R_z(\theta) L_z(\vartheta) u$ \cite{Ryblewski:2010ch}. This specific transformation is chosen in order to ensure that the 
four-vector $z^\mu$ has no transverse components in all frames. To find the necessary vectors 
in the lab frame based on the LRF expressions (\ref{eq:rfbasis}) we apply the inverse operation 
$X_{\alpha,\rm LAB}^\mu = (L_x R_z L_z)^{-1} X_{\alpha,\rm LRF}^\mu = (L_z)^{-1} (R_z)^{-1} (L_x)^{-1} X_{\alpha,\rm LRF}^\mu$ 
which is explicitly given by
\begin{equation}
X_{\alpha,\rm LAB}^\mu \! = 
\underbrace{\!\!
\left(
\begin{array}{cccc}
\cosh\vartheta & 0 & 0 & \sinh\vartheta\\
0 & 1 & 0 & 0 \\
0 & 0 & 1 & 0 \\
\sinh\vartheta & 0 & 0 & \cosh\vartheta 
\end{array}
\right)\!\!}_{(L_z)^{-1}} 
\underbrace{\!
\left(
\begin{array}{cccc}
1 & 0 & 0 & 0 \\
0 & \cos\phi & -\sin\phi & 0 \\
0 & \sin\phi & \cos\phi & 0 \\
0 & 0 & 0 & 1  
\end{array}
\right)\!\!}_{(R_z)^{-1}}  
\underbrace{\!
\left(
\begin{array}{cccc}
\cosh\psi  & \sinh\psi & 0 & 0 \\
\sinh\psi  & \cosh\psi & 0 & 0 \\
0 & 0 & 1 & 0 \\
0 & 0 & 0 & 1 
\end{array}
\right)\!\!}_{(L_x)^{-1}} \, X_{\alpha,\rm LRF}^\mu \, .
\end{equation}
which gives
\begin{equation}
\begin{array}{ll}
\begin{aligned}
u^0 &= \cosh\psi \cosh\vartheta \, , \\
u^1 &= \sinh\psi \cos\phi \, , \\
u^2 &= \sinh\psi \sin\phi \, , \\
u^3 &= \cosh\psi \sinh\vartheta \, ,
\end{aligned}
\quad \quad \quad &
\begin{aligned}
x^0 &= \sinh\psi \cosh\vartheta \, , \\
x^1 &= \cosh\psi \cos\phi \, , \\
x^2 &= \cosh\psi \sin\phi \, , \\
x^3 &= \sinh\psi \sinh\vartheta \, ,
\end{aligned}
\\
&\vspace{2mm}\\
\begin{aligned}
y^0 &= 0 \, , \\
y^1 &= -\sin\phi \, , \\
y^2 &= \cos\phi \, , \\
y^3 &= 0 \, ,
\end{aligned}
\quad \quad  \quad &
\begin{aligned}
z^0 &= \sinh\vartheta \, , \\
z^1 &= 0 \, , \\
z^2 &= 0 \, , \\
z^3 &= \cosh\vartheta \, .
\end{aligned}
\end{array}
\label{eq:expvectorbasis}
\end{equation}

In the limit that the system is boost invariant one can identify $\vartheta=\varsigma$, where $\varsigma$ is the spatial
rapidity defined through
\begin{eqnarray}
t&=&\tau \cosh\varsigma \, , \nonumber \\
z&=&\tau \sinh\varsigma \, ,
\label{eq:com-coord-1}
\end{eqnarray}
where $\tau = \sqrt{t^2 - z^2}$ is the proper time.  In the remainder of the paper when we
refer to a boost-invariant system we will use $\tau$ and $\varsigma$ as the longitudinal coordinates.


\subsection{Dynamical Equations}

In this section we derive the dynamical equations of motion by taking moments of
the Boltzmann equation \cite{GLW}
\begin{equation}
p^\mu \partial_\mu f(x,p) = - C[f] \, .
\label{eq:boltzmanneq1}
\end{equation}
The moments are defined by multiplying the left and right hand sides of the Boltzmann equation
by various powers of the four-momentum and then averaging in momentum space.  This can be
achieved via the $n^{\rm th}$ moment integral operator
\begin{equation}
{\hat{\cal I}}_n  \, \equiv \int d\chi \; p^{\mu_1} p^{\mu_2} \cdots p^{\mu_n}  \, ,
\end{equation}
where $n\geq 0$ is an integer and 
\begin{equation}
\int \! d\chi \equiv \int \! \! \frac{d^4{\bf p}}{(2\pi)^3} \, \delta(p_\mu p^\mu - m^2) \, 2 \theta(p^0) 
= \int \! \! \frac{d^3{\bf p}}{(2\pi)^3} \frac{1}{p^0} \, .
\end{equation}
%


\subsection{Zeroth moment of the Boltzmann equation}

The zeroth moment of the Boltzmann equation results from applying ${\hat{\cal I}}_0$ to both 
sides of (\ref{eq:boltzmanneq1})
\begin{eqnarray}
\int d\chi \; p^\mu \partial_\mu f 
&=& J_0 \, , 
\nonumber \\
\partial_\mu \int \! \! \frac{d^3{\bf p}}{(2\pi)^3} \frac{p^\mu}{p^0} \;  f 
&=& J_0 \, , 
\nonumber \\
\partial_\mu j^\mu
&=& J_0 \, ,
\end{eqnarray}
where $J_n \equiv - {\hat{\cal I}}_n C[f]$.  Note that we can rewrite the left hand side of the last expression as 
$j^\mu = n \;\! u^\mu$ where $n$ is the particle number density in the local rest frame.  Expanding we find
\begin{equation}
\partial_\mu j^\mu = D n + n \theta \, ,
\end{equation}
where 
\begin{eqnarray}
D &\equiv& u^\mu \partial_\mu \, , \nonumber \\
\theta &\equiv& \partial_\mu u^\mu \, ,
\label{eq:ddefs}
\end{eqnarray}
allowing us to write a general expression for the zeroth moment of the Boltzmann equation
\begin{equation}
D n + n \theta = J_0 \, .
\label{eq:zerothmomgen}
\end{equation}


\subsection{First moment of the Boltzmann equation}

The first moment of the Boltzmann equation is equivalent to the requirement of energy and momentum
conservation \cite{GLW}
\begin{equation}
\partial_\mu T^{\mu \nu} = 0 \, ,
\end{equation}
where $T^{\mu \nu}$ is the energy momentum tensor.  In the following we derive evolution equations
under different assumptions about the degree of symmetry of $T^{\mu \nu}$.


\subsubsection{Ideal hydrodynamics}

To begin we use the general form of the energy-momentum tensor for an isotropic system given
in Eq.~(\ref{eq:idealtmunu}) to obtain
\begin{equation}
\partial_\mu T^{\mu\nu} =  
u^\nu D ({\cal E} +{\cal P})   
+ u^\nu ({\cal E} +{\cal P}) \theta 
+ ({\cal E} +{\cal P}) D u^\nu  
- \partial^\nu {\cal P} \, ,
\end{equation}
where $D$ and $\theta$ are defined in Eq.~(\ref{eq:ddefs}).

Canonically one takes projections of $\partial_\mu T^{\mu\nu} = 0$ parallel and perpendicular to $u^\mu$.
The parallel projection is obtained via $u_\nu \partial_\mu T^{\mu\nu}$ which gives
\begin{eqnarray}
u_\nu \partial_\mu T^{\mu\nu} &=& D ({\cal E} +{\cal P})  
+ ({\cal E} +{\cal P}) \theta  
+ ({\cal E} +{\cal P}) u_\nu D u^\nu 
- D {\cal P} = 0
\nonumber \\
&=& D{\cal E} + ({\cal E} +{\cal P}) \theta = 0
\label{eq:idealeq1}
\end{eqnarray}
where we have used $u_\nu u^\nu = 1$ and  $u_\nu D u^\nu = \frac{1}{2} D(u_\nu u^\nu) = 0$.
This gives us our first equation for ideal hydrodynamics.  For the transverse projection we use 
$\Delta^{\mu\nu}$ defined in Eq.~(\ref{eq:transproj}) which satisfies $\Delta_{\alpha\nu} u^\nu = 0$.
This gives
\begin{equation}
{\Delta^\alpha}_\nu \partial_\mu T^{\mu\nu} = ({\cal E} +{\cal P}) {\Delta^\alpha}_\nu D u^\nu  
- {\Delta^\alpha}_\nu \partial^\nu {\cal P} = 0\, .
\end{equation}
Using the explicit form for ${\Delta^\alpha}_\nu = {g^\alpha}_\nu - u^\alpha u_\nu$ 
one obtains ${\Delta^\alpha}_\nu D u^\nu = D u^\alpha$.  We can additionally define
\begin{equation}
\nabla^\alpha \equiv {\Delta^\alpha}_\nu \partial^\nu = 
- \sum_{\beta=1}^3 X_\beta^\alpha X_{\nu\beta} \partial^\nu  \, ,
\end{equation}
which is the gradient in the spacelike directions.  Putting this together with Eq.~(\ref{eq:idealeq1}) one 
obtains the following two equations
\begin{eqnarray}
D{\cal E} + ({\cal E} +{\cal P}) \theta &=& 0 \, , \nonumber \\
({\cal E} +{\cal P}) D u^\alpha - \nabla^\alpha {\cal P} &=& 0 \, .
\label{eq:idealeq2}
\end{eqnarray}
In the second case $\alpha$ should be a spacelike index such that
we have four equations in total which should be supplemented by the equation of state 
which can be expressed in the form of a constraint on the trace of the energy momentum tensor 
${T^\mu}_\mu={\cal T}={\cal E} - 3{\cal P}$.


\subsubsection{Ideal Boost Invariant Dynamics with Transverse Expansion}

In this section we briefly review what happens when the system is boost invariant and we allow 
for inhomogeneities and flow in the transverse direction.  In this case we have from (\ref{eq:expvectorbasis})
\begin{equation}
u^\mu = (\cosh\psi \cosh\varsigma,\sinh\psi \cos\phi,\sinh\psi \sin\phi, \cosh\psi\sinh\varsigma) \, .
\end{equation}
It is convenient at this point to relabel the components of $u^\mu$ as
\begin{equation}
u^\mu = (u_0 \cosh\varsigma,u_x,u_y, u_0 \sinh\varsigma) \, .
\end{equation}
where the constraint $u_0^2 = 1 + u_x^2 + u_y^2$ should be satisfied.
Changing to proper time and spatial rapidity we obtain $u_\tau = u_0$, $u_\varsigma = 0$, and we have
\begin{eqnarray}
D &=& u^\mu \partial_\mu = u_0 \partial_\tau  + {\bf u}_\perp\!\cdot\nabla_\perp \, , \nonumber \\
\theta &=& \partial_\mu u^\mu = \partial_\tau u_0 + \nabla_\perp\!\cdot{\bf u}_\perp 
+ \frac{u_0}{\tau} \, .
\label{eq:ddefs3}
\end{eqnarray}
For the transverse gradient it is convenient to rewrite
\begin{equation}
\nabla^i = {\Delta^i}_\nu \partial^\nu = ({g^i}_\nu - u^i u_\nu) \partial^\nu = \partial^i - u^i D \, ,
\end{equation}
such that the second equation in (\ref{eq:idealeq2}) can be expanded into three equations
\begin{eqnarray}
({\cal E} + {\cal P}) D u_x + u_x D {\cal P} + \partial_x {\cal P} &= 0 \, , \nonumber \\
({\cal E} + {\cal P}) D u_y + u_y D {\cal P} + \partial_y {\cal P} &= 0 \, , \nonumber \\
({\cal E} + {\cal P}) D u_0 + u_0 D {\cal P} - \partial_\tau {\cal P} &= 0 \, ,
\label{eq:ide1}
\end{eqnarray}
which together with
\begin{equation}
D{\cal E} + ({\cal E} + {\cal P}) \theta = 0 \, ,
\end{equation}
would seem to give four equations for our four unknowns (${\cal E}$, ${\cal P}$, $u_x$, and 
$u_y$ since $u_0^2 = 1 + u_x^2 + u_y^2$); however, upon inspection one finds that 
Eqs.~(\ref{eq:ide1}) are not independent since $u_0$ times
the third equation is equal to $u_x$ times the first plus $u_y$ times the second.  We, 
therefore, have a choice of which equations to use and one can pick two of the three equations 
from (\ref{eq:ide1}), e.g. the first two.  The final equation is then provided canonically by
the equation of state which specifies, e.g., the energy density as a function of the pressure.


\subsection{Azimuthally-Symmetric Anisotropic Hydrodynamics}

We now proceed to the derivation of the dynamical equation for azimuthally-symmetric anisotropic hydrodynamics.
We remind the reader ``azimuthally-symmetric'' means that the momentum-space particle distribution
is azimuthally symmetric while the rotational symmetry in the $p_\perp$-$p_L$ plane is broken.
To begin we use the general form of the energy-momentum tensor for an azimuthally-symmetric anisotropic system given in Eq.~(\ref{eq:speroidaltmunu}) to obtain
\begin{eqnarray}
\partial_\mu T^{\mu\nu} &=&  
u^\nu D ({\cal E} +{\cal P}_\perp)   
+ u^\nu ({\cal E} +{\cal P}_\perp) \theta 
+ ({\cal E} +{\cal P}_\perp) D u^\nu  
- \partial^\nu {\cal P}_\perp  \nonumber \\
&&
\hspace{5mm}
+ z^\nu D_L ({\cal P}_L - {\cal P}_\perp)  
+ z^\nu ({\cal P}_L - {\cal P}_\perp) \theta_L 
+ ({\cal P}_L - {\cal P}_\perp)  D_L z^\nu = 0
\, ,
\end{eqnarray}
where 
\begin{eqnarray}
D_L &\equiv& z^\mu \partial_\mu \, , \nonumber \\
\theta_L &\equiv& \partial_\mu z^\mu \, .
\label{eq:ddefsspheroidal}
\end{eqnarray}

As before we take projections of $\partial_\mu T^{\mu\nu} = 0$ parallel and perpendicular to $u^\mu$.
The parallel projection is obtained via $u_\nu \partial_\mu T^{\mu\nu}$ which gives
\begin{equation}
u_\nu \partial_\mu T^{\mu\nu} = D {\cal E}
+ ({\cal E} +{\cal P}_\perp) \theta + ({\cal P}_L - {\cal P}_\perp)  u_\nu D_L z^\nu = 0 \, ,
\label{eq:speroidaleq1}
\end{equation}
where we have used $u_\nu u^\nu = 1$, $u_\nu D u^\nu = \frac{1}{2} 
D(u_\nu u^\nu) = 0$, and $u_\nu z^\nu = 0$.  This gives us our first equation 
for azimuthally-symmetric anisotropic hydrodynamics.  

For the transverse projection we use 
$\Delta^{\mu\nu}$ defined in Eq.~(\ref{eq:transproj}) which satisfies $\Delta_{\alpha\nu} u^\nu = 0$
and $\Delta_{\alpha\nu} z^\nu = z^\alpha$.
This gives
\begin{eqnarray}
{\Delta^\alpha}_\nu \partial_\mu T^{\mu\nu} 
 &=& ({\cal E} +{\cal P}_\perp) D u^\alpha - {\nabla}^\alpha {\cal P}_\perp 
 + z^\alpha D_L ({\cal P}_L - {\cal P}_\perp) + z^\alpha ({\cal P}_L - {\cal P}_\perp) \theta_L
 \nonumber \\
&&
\hspace{5mm}
+ ({\cal P}_L - {\cal P}_\perp)  D_L z^\alpha 
- ({\cal P}_L - {\cal P}_\perp)  u^\alpha u_\nu D_L z^\nu
= 0 \, .
\end{eqnarray}


\subsubsection{Boost Invariant Dynamics with Transverse Expansion}

In this case we have $z^\tau = 0$ and $z^\eta = 1/\tau$ such that
\begin{eqnarray}
D_L &=& z^\mu \partial_\mu = \frac{\partial_\varsigma}{\tau} \, , \nonumber \\
\theta_L &=& \partial_\mu z^\mu = 0 \, .
\label{eq:dldefs}
\end{eqnarray}
From the first line above we find $ u_\nu D_L z^\nu = u_0/\tau$.  This allows us
to simplify the parallel projection to
\begin{equation}
D {\cal E} + ({\cal E} +{\cal P}_\perp) \theta + ({\cal P}_L - {\cal P}_\perp)\frac{u_0}{\tau}  = 0 \, .
\label{eq:speroidaleq2}
\end{equation}
The transverse projections can also be simplified to
\begin{equation}
({\cal E} +{\cal P}_\perp) D u^\alpha + u^\alpha D {\cal P}_\perp + \partial_\alpha {\cal P}_\perp 
+ ({\cal P}_L - {\cal P}_\perp)  \left( \frac{\partial_\varsigma z^\alpha}{\tau} 
- \frac{u_0}{\tau}  u^\alpha \right)
= 0 \, ,
\end{equation}
from which we can then obtain three equations
\begin{eqnarray}
({\cal E} +{\cal P}_\perp) D u_x + u_x D {\cal P}_\perp + \partial_x {\cal P}_\perp 
+ ({\cal P}_\perp - {\cal P}_L) \frac{u_0 u_x}{\tau} &=& 0 \, , \nonumber \\
({\cal E} +{\cal P}_\perp) D u_y + u_y D {\cal P}_\perp + \partial_y {\cal P}_\perp 
+ ({\cal P}_\perp - {\cal P}_L) \frac{u_0 u_y}{\tau} &=& 0 \, , \nonumber \\
({\cal E} +{\cal P}_\perp) D u_0 + u_0 D {\cal P}_\perp - \partial_\tau {\cal P}_\perp 
+ ({\cal P}_\perp - {\cal P}_L) \frac{u_\perp^2}{\tau} &=& 0 \, . 
\end{eqnarray}
As was the case with ideal hydrodynamics, we see that $u_0$ times
the third equation is equal to $u_x$ times the first plus $u_y$ times the second
so that it is redundant.  This leaves us with the following three equations
\begin{eqnarray}
D {\cal E} + ({\cal E} +{\cal P}_\perp) \theta 
+ ({\cal P}_L - {\cal P}_\perp)\frac{u_0}{\tau}  &=& 0 \, , \nonumber \\
({\cal E} +{\cal P}_\perp) D u_x + \partial_x {\cal P}_\perp + u_x D {\cal P}_\perp
+ ({\cal P}_\perp - {\cal P}_L) \frac{u_0 u_x}{\tau} &=& 0 \, , \nonumber \\
({\cal E} +{\cal P}_\perp) D u_y + \partial_y {\cal P}_\perp + u_y D {\cal P}_\perp
+ ({\cal P}_\perp - {\cal P}_L) \frac{u_0 u_y}{\tau} &=& 0 \, . 
\end{eqnarray}
%

\subsection{Distribution function for azimuthally-symmetric systems}

We next consider the one-particle distribution function $f$ in the local rest frame and
show that in the case of a system that is locally azimuthally-symmetric in momentum space that it 
suffices to introduce one anisotropy parameter $\xi$
and a single scale $\Lambda$ \cite{Romatschke:2003ms}.  To begin we consider the general form
\begin{equation}
f(t,{\bf x},{\bf p}) = f_{\rm iso}(\sqrt{\bar{p}_\mu \Xi^{\mu\nu}(t,{\bf x}) \bar{p}_\nu}) \, .
\label{eq:rsformbreakdown}
\end{equation}
$\Xi^{\mu\nu}(t,{\bf x}) $ is a symmetric tensor, $f_{\rm iso}$ is an arbitrary isotropic distribution function, and 
${\bar p}^\mu \equiv p^\mu/\Lambda$, where $\Lambda(t,{\bf x})$ 
is a momentum scale that can depend on space and time (the so-called hard momentum scale).  
In the case  where the system is in thermal equilibrium, then
$f_{\rm iso}$ would be given by a Bose-Einstein or Fermi-Dirac distribution function. Note that the 
argument of the square root in $f_{\rm iso}$ 
should remain greater than or equal to zero in order for $f$ to be a single-valued real function.

If $\Xi^{\mu\nu}$ is a symmetric tensor and is diagonal in the local rest frame, we have
\begin{equation}
\Xi^{\mu\nu} = c_{00} u^\mu u^\nu  + \sum_{i=1}^3 c_{ii} X^\mu_i X^\nu_i  \, ,
\label{eq:Xitensorgen1}
\end{equation}
and if, additionally, the system is symmetric under  $x \leftrightarrow y$ then 
$c_{11} = c_{22} \equiv c_{\perp\perp}$ and we have 
\begin{eqnarray}
\Xi^{\mu\nu} &=& c_{00} u^\mu u^\nu  
+ c_{\perp\perp} \sum_{i=1}^2 X^\mu_i X^\nu_i  
+ c_{33} X^\mu_3 X^\nu_3 \, , \nonumber \\
&=& c_{00} u^\mu u^\nu  
- c_{\perp\perp} \Delta^{\mu\nu} 
+ (c_{33}-c_{\perp\perp}) X^\mu_3 X^\nu_3 \, .
\label{eq:Xitensorgen2}
\end{eqnarray}
Using our ability to redefine $\Lambda \rightarrow \sqrt{c_{00}} \Lambda$ in 
Eq.~(\ref{eq:rsformbreakdown}) we can rescale our coefficients.  Defining 
$c_{\perp\perp}/c_{00} \equiv \Phi$ and $(c_{33}-c_{\perp\perp})/c_{00} \equiv \alpha$ 
we can write compactly
\begin{equation}
\Xi^{\mu\nu} = u^\mu u^\nu - \Phi \Delta^{\mu\nu} + \alpha z^\mu z^\nu \, .
\end{equation}
Contracting with four-momenta on both sides we find
\begin{eqnarray}
p_\mu \Xi^{\mu \nu} p_\nu &=& p_0^2 + \Phi {\bf p}^2 + \alpha p_z^2 \, , 
\nonumber \\
&=& m^2 + (1+\Phi) {\bf p}^2 + \alpha p_z^2 \, ,
\end{eqnarray}
where we have used $p_0^2 = {\bf p}^2 + m^2$.  If we have a system of massless particles then
\begin{equation}
p_\mu \Xi^{\mu \nu} p_\nu = (1+\Phi)p_\perp^2 + (1 + \Phi + \alpha) p_z^2 \, ,
\end{equation}
and in this case we can once again use our ability to rescale $\Lambda \rightarrow \sqrt{(1+\Phi)} \Lambda$ 
and defining $1+\xi \equiv (1 + \Phi + \alpha)/(1 + \Phi)$ we obtain
\begin{equation}
p_\mu \Xi^{\mu \nu} p_\nu = p_\perp^2 + (1 + \xi) p_z^2 \, ,
\end{equation}
which has the form of the argument of the original one-dimensional Romatschke-Strickland (RS) distribution function 
\cite{Romatschke:2003ms}.

\subsection{Number density and Energy-Momentum Tensor with the RS distribution function}

Based on the results of the last section, the functional form of the RS distribution function
for a locally azimuthally-symmetric expanding anisotropic plasma is
\begin{equation}
\label{eq:rsansatz-long}
f({\bf x},{\bf p},\tau)= f_{\rm RS}({\bf p},\xi,\Lambda)= f_{\rm iso}\bigl(\sqrt{[{\bf p}_\perp^2+(1+\xi)p^2_z]/\Lambda^2}\bigr) \, ,
\end{equation}
where it is understood that on the right hand side $\xi$ and $\Lambda$ can depend on space and time.
Using this distribution function the number density is given by \cite{Romatschke:2004jh,%
Martinez:2009ry}
\begin{equation}
n(\xi,\Lambda) = \int \frac{d^3{\bf p}}{(2 \pi)^3} f _{\rm RS} = \frac{n_{\rm iso}(\Lambda)}{\sqrt{1+\xi}} \, .
\label{eq:nrs}
\end{equation}
where $n_{\rm iso}(\Lambda)$ is the number density one obtains in the isotropic limit.
 
One can also evaluate the energy-momentum tensor in the LRF 
\begin{equation}
T^{\mu\nu}= \int \frac{d^3{\bf p}}{(2 \pi)^3} \frac{p^\mu p^\nu}{p_0} f(\tau,{\bf x},{\bf p}) \, .
\end{equation}
By using the RS form (\ref{eq:rsansatz-long}) one gets the explicit components of the energy-momentum tensor~\cite{Martinez:2009ry}
\begin{subequations}
\label{eq:momentsanisotropic}
\begin{align}
\label{eq:energyaniso}
{\cal E}(\Lambda,\xi) &= T^{\tau\tau} = {\cal R}(\xi)\,{\cal E}_{\rm iso}(\Lambda)\, ,\\
\label{eq:transpressaniso}
{\cal P}_\perp(\Lambda,\xi) &= \frac{1}{2}\left( T^{xx} + T^{yy}\right) = {\cal R}_\perp(\xi){\cal P}_{\rm iso}(\Lambda)\, , \\
\label{eq:longpressaniso}
{\cal P}_L(\Lambda,\xi) &= - T^{\varsigma}_\varsigma = {\cal R}_{\rm L}(\xi){\cal P}_{\rm iso}(\Lambda)\, ,
\end{align}
\end{subequations}
where ${\cal P}_{\rm iso}(\Lambda)$ and ${\cal E}_{\rm iso}(\Lambda)$ are the isotropic 
pressure and energy density, respectively, and  
\begin{subequations}
\begin{align}
{\cal R}(\xi) &\equiv \frac{1}{2}\left(\frac{1}{1+\xi} +\frac{\arctan\sqrt{\xi}}{\sqrt{\xi}} \right) \, , \\
{\cal R}_\perp(\xi)  &\equiv \frac{3}{2 \xi} \left( \frac{1+(\xi^2-1){\cal R}(\xi)}{\xi + 1}\right) \, , \\
{\cal R}_L(\xi) &\equiv \frac{3}{\xi} \left( \frac{(\xi+1){\cal R}(\xi)-1}{\xi+1}\right) \, .
\end{align}
\label{eq:rfuncs}
\end{subequations}

The equation of state can be imposed as a relationship
between $\eiso$ and $\piso$.  In what follows we will assume an ideal equation of state which
is appropriate for a conformal massless gas, i.e. $\eiso = 3 \piso$.

\subsection{Relaxation time approximation}

As mentioned in previous sections the dynamical equations necessary can be obtained by taking moments
of the Boltzmann equation $p^\mu \partial_\mu f = - C[f]$.  Here we use the relaxation time approximation
with relaxation rate $\Gamma$
\begin{equation}
{\cal C}[f_{RS}] = p_\mu u^\mu \, \Gamma \, \left[ f_{\rm RS}({\bf p},\xi,\Lambda,\varsigma) - f_{\rm eq}(|{\bf p}|,T) \right]\,,
\label{rel-time}
\end{equation}
where $\varsigma$ is the spatial rapidity and we fix $\Gamma$ such that the 2nd-order viscous 
hydrodynamical equations are reproduced in the one-dimensional
transversally symmetric case \cite{Martinez:2010sc}.  This requires that
\begin{eqnarray}
\Gamma &\equiv& \frac{2}{\tau_\pi}\,, \nonumber \\ 
\tau_\pi &\equiv& \frac{5}{4}\frac{\eta}{\cal P}\, ,
\end{eqnarray}
\label{eq:hydromatch}
which for an ideal equation of state results in
\begin{equation}
\Gamma = \frac{2T(\tau)}{5\bar\eta} = \frac{2{\cal R}^{1/4}(\xi)\Lambda}{5\bar\eta} \, ,
\label{eq:gammamatch}
\end{equation}
where $\bar\eta = \eta/{\cal S}$ with $\eta$ being the shear viscosity and ${\cal S}$ being
the entropy density.  
We note that one could perform a matching to 2nd-order viscous hydrodynamics including transverse 
dynamics, but we have not attempted to do so.  Instead we use the 1d matching above and in the 
results section we show that numerical results from viscous hydrodynamics codes which include 
transverse dynamics are reproduced for small $\bar\eta$.  That being said, we have no reason to 
expect that the linearized equations would not reproduce 2nd-order viscous hydrodynamics; however, 
this remains to be proven.

\subsection{Dynamical Equations of Motion}

Based on the results of the previous sections, we can derive the explicit form of the dynamical 
equations of motion for a (2+1)-dimensional boost invariant system.  

\subsubsection{Zeroth moment of the Boltzmann Equation}

For the RS form the 0th moment of the Boltzmann equation (\ref{eq:zerothmomgen}) is written as

\begin{equation}
\frac{1}{1+\xi}D\xi - 6D(\log\Lambda) - 2 \theta = 2 \Gamma\left(1 - {\cal R}^{3/4}(\xi) \sqrt{1+\xi}\right) \, .
\label{eq:zeromom}
\end{equation}
where we used explicitly the functional form of particle density $n$~(\ref{eq:nrs}) and the scattering kernel for relaxation time approximation~(\ref{rel-time}). 

\subsubsection{First moment of the Boltzmann Equation}

Using the RS form one finds the following three equations by requiring energy-momentum conservation
\begin{eqnarray}
{\cal R}'(\xi) D\xi + 4 {\cal R}(\xi) D(\log\Lambda) &=& 
- \left({\cal R}(\xi) + \frac{1}{3} {\cal R}_\perp(\xi)\right) \Delta_\perp
- \left({\cal R}(\xi) + \frac{1}{3} {\cal R}_L(\xi)\right) \frac{u_0}{\tau} \, ,
\nonumber \\
\left[3{\cal R}(\xi) + {\cal R}_\perp(\xi)\right] D u_\perp &=&-u_\perp \left[ {\cal R}_\perp'(\xi) \tilde{D} \xi 
+ 4  {\cal R}_\perp(\xi) \tilde{D} (\log\Lambda) + \frac{u_0}{\tau} ({\cal R}_\perp(\xi)-{\cal R}_L(\xi)) \right] ,
\nonumber \\
u_y^2 \left[3{\cal R}(\xi) + {\cal R}_\perp(\xi)\right] D \left( \frac{u_x}{u_y} \right) &=& 
{\cal R}_\perp'(\xi) D_\perp\xi 
+ 4 {\cal R}_\perp(\xi) D_\perp(\log\Lambda) \, ,
\label{eq:firstmom}
\end{eqnarray}
where
\begin{eqnarray}
\Delta_\perp &\equiv& \partial_\tau u_0 + \nabla_\perp \cdot {\bf u}_\perp \, , \nonumber \\
\tilde{D} &\equiv& u_0 \partial_\tau + \frac{u_0^2}{u_\perp^2} {\bf u}_\perp \cdot \nabla_\perp \, , \nonumber \\
D_\perp &\equiv& \hat{\bf z} \cdot ({\bf u}_\perp \times \nabla_T) = u_x \partial_y - u_y \partial_x \, ,
\end{eqnarray}
${\bf u}_\perp  \equiv (u_x,u_y)$, and $u_0^2 = 1 + u_\perp^2$.

\section{Initial Conditions}
\label{sec:initialconditions}

We consider collisions of symmetric nuclei, each containing $A$ nucleons.  We will
study both participant and binary collision type initial conditions \cite{Bialas:1976ed} using a Woods-Saxon 
distribution for each nuclei's transverse profile \cite{Glauber:1970jm}.  For an individual nucleus we 
take the density to be
\begin{equation}
n_A(r) = \frac{n_0}{1 + e^{(r-R)/d}} \, ,
\end{equation}
where $n_0 = 0.17\;{\rm fm}^{-3}$ is the central nucleon 
density, $R = (1.12 A^{1/3} - 0.86 A^{-1/3})\;{\rm fm}$ 
is the nuclear radius, and $d = 0.54\;{\rm fm}$ is the ``skin depth''.  The density is normalized such that 
$\lim_{A\to\infty} \int d^3r \, n_A(r) = A$, where $A$ is the total number of nucleons in the nucleus.  
The normalization condition fixes $n_0$ to the value specified above.
From the nucleon
density we first construct the thickness function in the standard way by integrating over the longitudinal 
direction, i.e.
\begin{equation}
T_A(x,y) = \int_{-\infty}^{\infty} dz \, n_A(\sqrt{x^2+y^2+z^2}) \, .  
\end{equation}
With this in hand we can construct the  overlap density between two nuclei whose centers are 
separated by an impact parameter vector $\vec{b}$ which we choose to point along the $\hat{x}$ 
direction, i.e. $\vec{b} = b \hat{x}$.  We choose to locate the origin of our coordinate
system to lie halfway between the center of the two nuclei such that the overlap density can be
written as
\begin{equation}
n_{AB}(x,y,b) =  T_A(x+b/2,y) T_B(x-b/2,y) \, .
\label{eq:nab}
\end{equation}
Another quantity of interest is the participant density which is given by 
\begin{eqnarray}
n_{\rm part}(x,y,b) &=& T_A(x+b/2,y) \left[ 1- \left(1-\frac{\sigma_{NN}\,T_B(x-b/2,y)}{B}\right)^{\!\!B} \right]
\nonumber \\
&& \hspace{2cm} + \; T_B(x-b/2,y) \left[ 1- \left(1-\frac{\sigma_{NN}\,T_A(x+b/2,y)}{A}\right)^{\!\!A} \right] \, .
\end{eqnarray}
For LHC collisions at $\sqrt{s_{NN}}=2.76$ TeV we use $\sigma_{NN}$ = 62 mb and for RHIC
collisions at $\sqrt{s_{NN}}=200$ GeV we use $\sigma_{NN}$ = 42 mb.
From the participant density we construct our first possible initial condition for the 
transverse energy density profile at central rapidity
\begin{equation}
{\cal E}_0^{\rm part} = {\cal E}_0 \, \frac{n_{\rm part}(x,y,b)}{n_{\rm part}(0,0,0)} \, ,
\end{equation}
where ${\cal E}_0$ is the central energy density obtained in a central collision between the two nuclei.

As an alternative initial condition for energy density one could use the number of binary collisions 
which is defined as
\begin{equation}
n_{\rm coll}(x,y,b) = \sigma_{NN} \, n_{AB}(x,y,b)\, .
\end{equation}
from which we obtain the binary collision energy scaling
\begin{equation}
{\cal E}_0^{\rm coll} = {\cal E}_0 \, \frac{n_{\rm coll}(x,y,b)}{n_{\rm coll}(0,0,0)} 
= {\cal E}_0 \, \frac{n_{AB}(x,y,b)}{n_{AB}(0,0,0)} \, .
\end{equation}

\section{Numerical Methods}
\label{sec:nummeth}

We consider both smooth and fluctuating initial conditions using three numerical algorithms.  In the following
two subsections we describe the implementation of each algorithm.  In each case detailed below the code is implemented
using the C programming language.

\subsection{Centered Differences Algorithm}

In the first algorithm which we will refer to as the ``centered-differences algorithm'' we solve Eqs.~(\ref{eq:zeromom}) 
and Eqs.~(\ref{eq:firstmom}) by first analytically solving for the individual proper-time derivatives of the four dynamical 
variables: $\xi$, $\Lambda$, $u_x$, and $u_y$ using \textsc{Mathematica} \cite{mathematicaV7}.  
We then had \textsc{Mathematica} output, in C format, the 
necessary right hand sides of the four update equations.  We then discretize space on a regular square lattice with
lattice spacing, $\Delta x = a$.  For the spatial derivatives we use centered differences except on the edges of the lattice where 
we apply either a left- or right-handed first order derivative.  For the temporal
updates we use fourth-order Runge-Kutta (RK4) with a step size of $\Delta t = \epsilon$.

For smooth initial conditions the previous method suffices; however, for fluctuating initial conditions one finds that
using centered differences introduces spurious oscillations in regions where there are large gradients.  In order to damp
these oscillations one could attempt to use a two-dimensional Lax-Friedrichs (LAX) update 
\cite{Lax1954,Friedrichs1954}.  In practice this amounts to replacing the current value of a given dynamical variable 
by a local spatial average over neighboring sites and using this as a stand in for the current value of the variable, e.g.
\begin{equation}
\xi_{\rm LAX}(\tau,x,y) = \left[ \xi(\tau,x+a,y)+\xi(\tau,x-a,y)+\xi(\tau,x,y+a)+\xi(\tau,y-a) \right] /4,
\label{eq:fLAX}
\end{equation}
and now the $\xi$ update for a temporal step of size $\epsilon$ becomes schematically
\begin{equation}
\xi(\tau+\epsilon,x,y) = \xi_{\rm LAX}(\tau,x,y) + \epsilon \, {\rm RHS}_\xi(\tau,x,y) \, ,
\end{equation}
where ${\rm RHS}_\xi$ stands for the (rather complicated) right hand size of the $\xi$ update equation.  However,
such a scheme results in too much numerical dissipation.  An alternative is to realize that the source of the spurious
oscillations is the weak coupling between odd- and even-number lattice sites.  The full LAX scheme above maximally
couples these
interleaving lattices; however, this need not be done.  Instead one can weight the LAX-smoothed values with a weight
$\lambda$ and combine this with the current value of the variable in question, e.g.
\begin{equation}
\xi_{\rm wLAX}(\tau,x,y) = \lambda\xi_{\rm LAX}(\tau,x,y) + (1-\lambda) \xi(\tau,x,y) \, .
\label{eq:wLAX}
\end{equation}
The smaller the value of $\lambda$, the less the numerical viscosity.
In practice, we have found that for the \ahydro equations one should take $\lambda >0.02$ in order to achieve
numerical stability.  In the results section below we use $\lambda = 0.05$ which represents a factor of twenty
decrease in the dissipation induced by LAX-smoothing.  Note that,
when activated, wLAX smoothing is implemented for all dynamical variables ($\Lambda$, $\xi$, $u_x$, and $u_y$)
after each full time step of $\epsilon$ and
not within each RK4 substep.  We will only need to use the wLAX method for fluctuating initial conditions; however,
in App.~\ref{sec:numtests} 
we present numerical tests using it in the smooth initial conditions case in order to show that the amount
of numerical viscosity in the wLAX case is not numerically significant.  That being said, one would also like to have
another method for handling the spurious oscillations caused by using higher-order centered differences.  This has
motivated us to also implement the Kurganov-Tadmor central scheme which we describe in the next subsection.

\subsection{The MUSCL Algorithm}

As mentioned above, when there are large gradients present in a hyperbolic partial differential equation, the application
of straightforward centered-differences scheme can lead to spurious oscillations.
For smooth initial conditions and finite shear viscosity this is not an issue; however, for fluctuating initial conditions
one needs a way to handle shocks and discontinuities.
One way to proceed is to implement the LAX method as described previously; however, the LAX method introduces
numerical viscosity into the algorithm which scales like the $(\Delta x)^2/\Delta t$ so that it is not possible
to take the temporal step size to zero without having extremely small lattice spacing to reduce the numerical viscosity.
As discussed above one can reduce the amount of numerical viscosity by instead using the weighted LAX (wLAX) prescription
described above; however, it is desirable to have an alternative algorithm in order to be sure of the results.

For this purpose we have also implemented a ``Monotone Upstream-Centered Schemes for Conservation Laws'' 
(MUSCL) scheme derived by Kurganov and Tadmor \cite{KurganovTadmor2000} which has been extended to include 
nonlinear sources \cite{NaidooBaboolal2004}.
This method is particularly appealing because it can be shown that, although it does induce some numerical viscosity, the
magnitude of the numerical viscosity induced scales like as a power of the lattice spacing with no power of the temporal step 
size in the denominator allowing one to take extremely small time steps without inducing large artificial numerical 
viscosity.
Our implementation closely follows that introduced by Schenke et al. \cite{Schenke:2010nt} to solve three-dimensional
relativistic ideal hydrodynamics equations.  They have also extended the method to 2nd-order three-dimensional relativistic
viscous hydrodynamics \cite{Schenke:2010rr,Schenke:2011tv} with fluctuating initial conditions.

To explain the algorithm let us consider the simpler case of a one dimensional system of hyperbolic partial differential 
equations which can be cast into ``conservative'' form, i.e.
\begin{equation}
\partial_t u + F_x(u) = 0 \, ,
\label{eq:consform1} 
\end{equation}
where $u$ is, in general, an n-dimensional vector, $F$ is a so-called flux variable or flux function,
and $F_x(u) = \partial_x F(u)$.  
For example, if one were solving the advection equation, $\partial_t u 
+ \partial_x u= 0$ then we would have $F = u$ and if one were solving Burgers' equation 
$\partial_t u + u \partial_x u = 0$ this can be written in conservative form as 
$\partial_t u + \partial_x (u^2/2) = 0$ so that, in this case, $F = u^2/2$. 
Given a partial differential equation of the form (\ref{eq:consform1})
Kurganov and Tadmor derived the following semi-discrete update equation
\begin{equation}
\frac{d u_j}{d t} = - \frac{H_{j+1/2}(t) - H_{j-1/2}(t)}{\Delta x} \, ,
\label{eq:ktupdate1}
\end{equation}
where the numerical flux function $H$ is given by
\begin{equation}
H_{j+1/2}(t) \equiv \frac{ F\left(u^+_{j+1/2}(t)\right) + F\left(u^-_{j+1/2}(t)\right) }{2}
	- \frac{a^x_{j+1/2}(t)}{2} \left[u^+_{j+1/2}(t) - u^-_{j+1/2}(t)\right] \, ,
\label{eq:numflux}
\end{equation}
with $a^x_{j+1/2}(t)$ being the local propagation velocity in the $x$-direction which is given by the 
maximum of the left and right half-site extrapolated spectral radius of $\partial F/\partial u$ which is
defined as $\rho$
\begin{equation}
a^x_{j+1/2}(t) \equiv {\rm max} \! \left\{  
\rho\!\left(\frac{\partial F}{\partial u}\left(u^+_{j+1/2}(t)\right) \right) , \,
\rho\!\left(\frac{\partial F}{\partial u}\left(u^-_{j+1/2}(t)\right) \right)
\right\} \, ,
\label{eq:propvel}
\end{equation}
and finally, the half-site extrapolated intermediate values $u^\pm_{j+1/2}$ are given by
\begin{eqnarray}
u^+_{j+1/2} &\equiv& u_{j+1}(t) - \frac{\Delta x}{2} (u_x)_{j+1}(t) \, , \nonumber \\
u^-_{j+1/2} &\equiv& u_j(t) + \frac{\Delta x}{2} (u_x)_j(t) \, .
\label{eq:intmedvals}
\end{eqnarray}
For the derivatives, $u_x$, appearing in (\ref{eq:intmedvals}) one should use a total variation diminishing
``flux-limiter'' so that spurious oscillators are avoided  \cite{Harten1983}.  
We follow the original
paper of Kurganov and Tadmor and use the three-argument minmod flux-limiter \cite{vanLeer1974}:
\begin{equation}
(u_x)_j = {\rm minmod}\!\left( 
\theta \frac{u_j - u_{j-1}}{\Delta x} , 
\frac{u_{j+1}- u_{j-1}}{2 \Delta x} ,
\frac{u_{j+1} - u_j}{\Delta x} \right), \;\;\; 1 \leq \theta \leq 2 \, ,
\end{equation}
where
\begin{equation}
\hbox{minmod}(x_1, x_2, \cdots) = \left\{
\begin{array}{ll}
\hbox{min}_j\{x_j\},& \hbox{if $x_j> 0$ $\forall \, j$}\\
\hbox{max}_j\{x_j\},& \hbox{if $x_j< 0$ $\forall \, j$}\\
0 & \hbox{otherwise} \;\;\;\;\;\; .
\end{array}
\right. 
\end{equation}
The value of $\theta$ controls the dissipation of the flux limiter with $\theta=1$ being the most dissipative
and $\theta=2$ being the least.  In this paper we follow \cite{Schenke:2010nt} and use $\theta = 1.1$.
For details of the derivation of the Kurganov-Tadmor scheme we refer the reader to their original paper 
\cite{KurganovTadmor2000}.  As mentioned above one can extend the Kurganov-Tadmor scheme to accommodate
nonlinear time-dependent sources.  Including the possibility of a time-dependent source
changes our one-dimensional example to
\begin{equation}
\partial_t u + F_x(u) = J(t,u) \, ,
\label{eq:consform2} 
\end{equation}
where $J$ is a source term.  Naidoo and Baboolal \cite{NaidooBaboolal2004} demonstrated that, in this case, 
only a simple modification of adding the source on the right hand side was necessary
\begin{equation}
\frac{d u_j}{d t} = - \frac{H_{j+1/2}(t) - H_{j-1/2}(t)}{\Delta x} + J(t,u_j) \, ,
\label{eq:ktupdate2source}
\end{equation}
We note that to extend the method described thus far to 
multiple dimensions one introduces flux functions for each direction, e.g. $F_y$ and $F_z$,
and includes these in the update rule by defining new numerical flux functions (\ref{eq:numflux}) and 
propagation velocities (\ref{eq:propvel}) accordingly.

\subsubsection{Applying MUSCL to \ahydro}

In the case of \ahydro all of the evolution equations stem from conservative systems with sources, therefore we
can apply the general method just described.  For this purpose we need the first and second moments of the Boltzmann
equation with the RS form for the one-particle distribution function.  The zeroth moment 
can be written in a conservative form with sources in $\tau$-$\varsigma$ coordinates as follows
\begin{equation}
\partial_\tau j^\tau + \nabla_\perp \cdot {\bf j}^\perp = -\frac{j^\tau}{\tau} + J_0 \, ,
\end{equation}
where $j^\mu = n \, u^\mu$ is the particle four-current and
\begin{equation}
J_0 \equiv \Gamma \, n_{\rm iso}(\Lambda) \left[ \frac{1}{\sqrt{1+\xi}} - {\cal R}^{3/4}(\xi) \right] \, ,
\end{equation}
is the zeroth-moment of the right-hand side of the Boltzmann equation in the relaxation time approximation
used herein.  The remaining three
update equations necessary can be obtained from energy-momentum conservation, $\partial_\mu T^{\mu \nu} =0$,
giving
\begin{eqnarray}
\partial_\tau T^{\tau \tau} + \partial_x T^{\tau x} + \partial_y T^{\tau y} 
&=&
-\frac{1}{\tau} \left[ T^{\tau\tau} + \tau^2 T^{\varsigma\varsigma}\right] \, , \\
\partial_\tau T^{\tau x} + \partial_x T^{x x} + \partial_y T^{x y} 
&=&
-\frac{T^{\tau x}}{\tau} \, , \\
\partial_\tau T^{\tau y} + \partial_x T^{x y} + \partial_y T^{y y} 
&=&
-\frac{T^{\tau y}}{\tau} \, .
\end{eqnarray}
Once the dynamical variables $j^\tau$, $T^{\tau \tau}$, $T^{\tau x}$, $T^{\tau y}$ are updated via
these equations, they can then can be used to construct the remaining components of $j^\mu$ and 
$T^{\mu \nu}$.  In our case it is necessary to solve
two simultaneous nonlinear equations for $\xi$ and $\Lambda$ which will then allow us to determine the rest
of the information necessary to proceed with the solution.  To see how this works in practice, we first use 
(\ref{eq:speroidaltmunu}) to write the non-vanishing components of $T^{\mu\nu}$ and 
$j^\mu = n \, u^\mu$ explicitly 
\begin{subequations}
\begin{eqnarray}
T^{\tau\tau} &=& ({\cal E}+{\cal P}_\perp) u^0 u^0 - {\cal P}_\perp \, , \\
T^{\tau i} &=& ({\cal E}+{\cal P}_\perp) u^0 u^i \, , \\
T^{ij} &=& ({\cal E}+{\cal P}_\perp) u^i u^j \, , \\
T^{ii} &=& ({\cal E}+{\cal P}_\perp) u^i u^i + {\cal P}_\perp \, , \\
T^{\varsigma\varsigma} &=& {\cal P}_L/\tau^2 \, , \\
j^\tau &=& n \, u^0 \, , \\
j^i &=& n \, u^i \, ,
\end{eqnarray}
\label{eq:explicittmunu}
\end{subequations}
where $i \in \{x,y\}$.
Using these equations and the normalization condition $u_\tau^2 = 1 + u_x^2 + u_y^2$ 
one finds two nonlinear equations, similar to those obtained in Ref.~\cite{Schenke:2010nt},
\begin{equation}
{\cal E}(\Lambda,\xi) = T^{\tau \tau} - \frac{ (T^{\tau x})^2 + (T^{\tau y})^2 }{T^{\tau \tau} 
+ {\cal P}_\perp(\Lambda,\xi)} \, ,
\label{eq:ktone}
\end{equation}
and
\begin{equation}
j^\tau = n(\Lambda,\xi) \left[ \frac{T^{\tau \tau} + {\cal P}_\perp(\Lambda,\xi)}{{\cal E}(\Lambda,\xi) 
+ {\cal P}_\perp(\Lambda,\xi)} \right] .
\label{eq:kttwo}
\end{equation}
From these two equations one can numerically solve for $\Lambda$ and $\xi$.  These values can then be 
used to determine $u^\tau$ and $u^i$ via 
\begin{subequations}
\begin{eqnarray}
u^\tau &=& \frac{j^\tau}{n(\Lambda,\xi)} \, , \\
u^i &=& \frac{n(\Lambda,\xi) \, T^{\tau i} }{j^\tau \, [{\cal E}(\Lambda,\xi)  + {\cal P}_\perp(\Lambda,\xi)]} \, .\end{eqnarray}
\end{subequations}
Once determined, these components of the four-velocity together with the values of $\Lambda$ and 
$\xi$ can be used to determine all remaining variables in (\ref{eq:explicittmunu}).

The only remaining ingredient necessary for the Kurganov-Tadmor algorithm to be implemented fully is
to determine the local propagation velocities $a^i_{j+1/2}(t)$.  These are obtained by evaluating
the eigenvalues of the $4\times4$ Jacobian of $j^\tau$, $T^{\tau \tau}$, $T^{\tau x}$, $T^{\tau y}$.
As was the case in Ref.~\cite{Schenke:2010nt}, with some work and a little bit of help from \textsc{Mathematica},
one finds that two of the four eigenvalues are degenerate and equal to $u^i/u^\tau$ and the other two are given by
\begin{equation}
\lambda^\pm_i = \frac{A\pm\sqrt{B}}{D} \, ,
\end{equation}
with
\begin{subequations}
\begin{eqnarray}
A &=& u^\tau u^i ( 1 - v^2 ) \, , \\
B &=& \left[u_\tau^2 - u_i^2 - (u_\tau^2-u_i^2-1) v^2 \right] v^2 \, , \\
D &=& u_\tau^2 - (u_\tau^2-1) v^2 \, ,
\end{eqnarray}
\end{subequations}
and
\begin{equation}
v^2 = \frac{\partial {\cal P}_\perp}{\partial {\cal E}} 
+ \frac{n}{{\cal E}+{\cal P}_\perp} \frac{\partial {\cal P}_\perp}{\partial n} \, .
\end{equation}
Using an ideal equation of state for which ${\cal E}_{\rm iso} = 3 {\cal P}_{\rm iso}$ one obtains
\begin{equation}
v^2(\xi) = \frac{1}{3} \frac{2 {\cal R}_\perp(\xi) + 3 (1+\xi) {\cal R}_\perp^\prime(\xi)}{2 {\cal R}(\xi) + 3 (1+\xi) {\cal R}^\prime(\xi)}
+ \frac{4(1+\xi)}{3{\cal R}(\xi) + {\cal R}_\perp(\xi)}
 \frac{{\cal R}^\prime(\xi) {\cal R}_\perp(\xi) - {\cal R}(\xi) {\cal R}_\perp^\prime(\xi)}{2 {\cal R}(\xi) + 3 (1+\xi) {\cal R}^\prime(\xi)} \, .
\end{equation}
In this function both terms individually diverge in the limit that $\xi \rightarrow 0$, however, these divergences
cancel to give a finite result of $\lim_{\xi \rightarrow 0} v^2 = 2/5$.  It has other limits of $\lim_{\xi \rightarrow -1} v^2 = 0$ and $\lim_{\xi \rightarrow \infty} v^2 = 1/2$.  Using the now known eigenvalues one finds that the maximum value
of the four eigenvalues is given by
\begin{equation}
\rho = |{\rm max}(\lambda_i)| = \frac{|A| + \sqrt{B}}{D} \, .
\end{equation}

Using the above scheme one can evolve the \ahydro system with fluctuating initial conditions; however, there is
a caveat, namely that the linearly interpolated intermediate values of $j^\tau$, $T^{\tau\tau}$, $T^{\tau x}$, 
and  $T^{\tau y}$ determined via (\ref{eq:intmedvals}) may not have real-valued solutions for $\Lambda$ and
$\xi$ using Eqs.~(\ref{eq:ktone}) and (\ref{eq:kttwo}).  In practice, we find that it is necessary to use extremely fine
lattices in order to ameliorate this problem.  Alternatively, we have found that instead of extrapolating the four
variables $j^\tau$, $T^{\tau\tau}$, $T^{\tau x}$, and  $T^{\tau y}$ to the half-sites, one can instead
extrapolate the current values of $\Lambda$ and $\xi$ to the half-sites for use in evaluating the flux functions.  In
addition, we have found that in practice it is necessary to use a ``hybrid'' algorithm in which the centered-differences
scheme described in the previous subsection is used as the initial guess for the nonlinear root finder which solves
Eqs.~(\ref{eq:ktone}) and (\ref{eq:kttwo}).  This is necessary, in particular, in regions where $\xi\simeq0$ since
Eqs.~(\ref{eq:ktone}) and (\ref{eq:kttwo}) have two solutions which become very close together causing the root
finder to oscillate between the two solutions.  The predicted value from the centered-differences scheme predicts
for the nonlinear root finder which solution to use in this case.  We will refer to this method as 
``Hybrid Kurganov-Tadmor''.

\section{Results}
\label{sec:results}

In this section we present results for the time evolution of the matter generated in heavy ion collisions at 
LHC energies using the \ahydro evolution equations (\ref{eq:zeromom}) and (\ref{eq:firstmom}).
For the results presented here we assume a ideal gas of quarks and gluons with $N_f=2$ so that there are $N_{\rm dof}
= 37$ degrees of freedom.  For our numerical tests and results we will concentrate on the spatial and momentum-space 
ellipticities, $\epsilon_x$ 
\begin{equation}
\epsilon_x= \frac{\eavg{y^2 - x^2}_{\cal E}}{\eavg{x^2 + y^2}_{\cal E}} \, ,
\end{equation}
and $\epsilon_p$ is defined in the lab frame via
\begin{equation}
\epsilon_p = \frac{\eavg{T^{xx} - T^{yy}}}{\eavg{T^{xx} + T^{yy}}} \, ,
\end{equation}
where $\eavg{\!x^2\!}_{\cal E}$ and $\eavg{\!y^2\!}_{\cal E}$ are the proper-time dependent average values of $x^2$ 
and $y^2$ weighted by the energy density 
\begin{equation}
\eavg{x^2}_{\cal E} \, \equiv \, {\cal N} \int_{x,y} x^2 {\cal E}(\tau,x,y) \, ,
\end{equation}
and the averages in the momentum-space ellipticity represent unweighted integrals over the transverse
directions.

Note that the normalization ${\cal N}$ is arbitrary since it cancels in the ratio we are computing.  These definitions
are the conventional ones from the literature \cite{Kolb:2000sd} which, unfortunately, are slightly inconsistent 
since $\epsilon_x$ is defined in the local rest frame and $\epsilon_p$ in the lab frame.  It would be more consistent 
to weight the spatial average by $T^{\tau\tau}$; however, to be consistent with the existing literature we will use 
the definition weighted with the energy density in the local rest frame.

\begin{figure}
\begin{center}
\includegraphics[width=8.1cm]{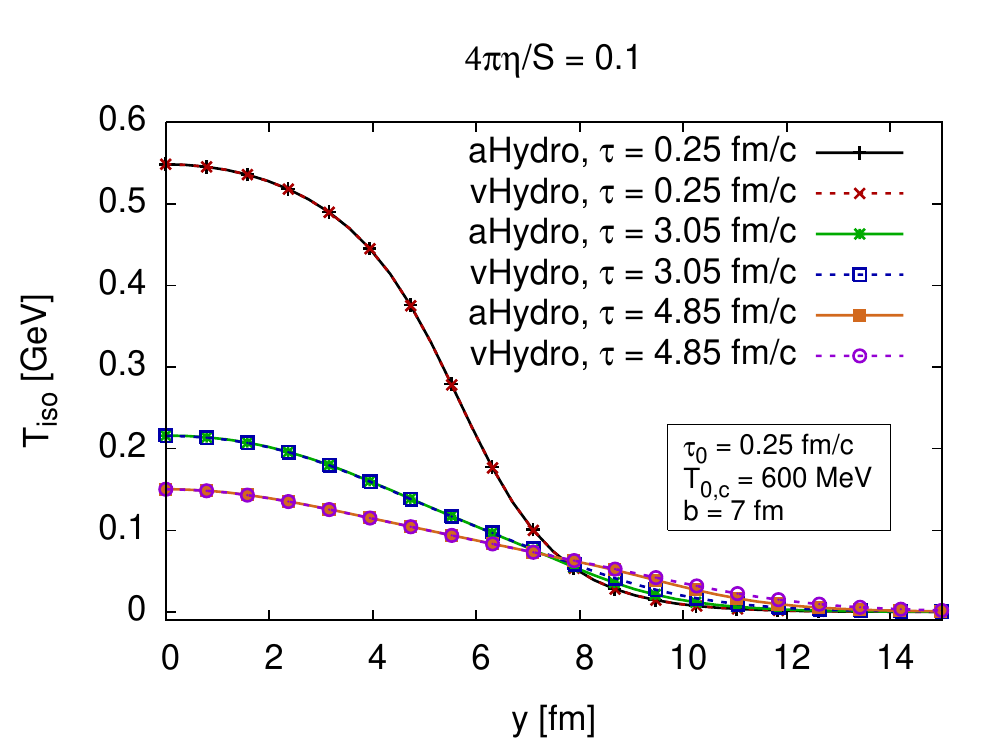}
\includegraphics[width=8.1cm]{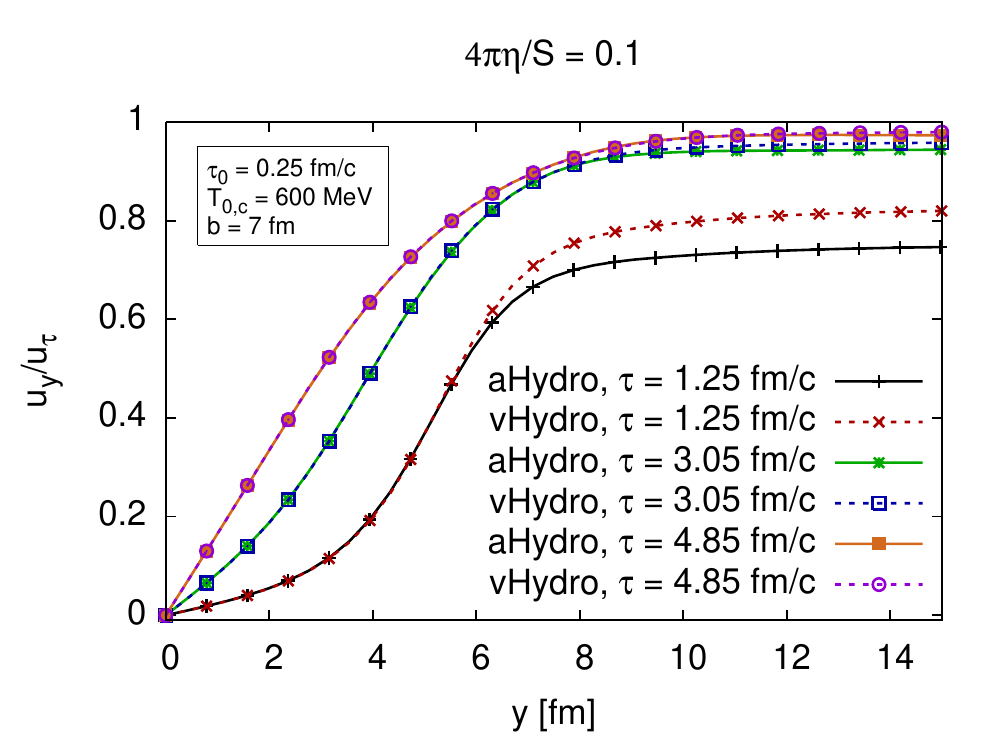}
\end{center}
\vspace{-8mm}
\caption{(Color online) Comparison of \ahydro isotropic temperature and flow profiles with 2nd-order viscous hydrodynamics code for
$4 \pi \eta/{\cal S}=0.1$ and $b=7$ fm.  Lattice size used was $109\times109$ with $a=0.394$ fm, $\epsilon=0.01$ fm/c, 
$\tau_0 = 0.25$ fm/c, $\Lambda_0 = $ 600 MeV, and $\xi_0 = 0$.  For the transverse profile Glauber binary collision scaling
was used.}
\label{fig:vCompare1}
\end{figure}

\begin{figure}
\begin{center}
\includegraphics[width=8.1cm]{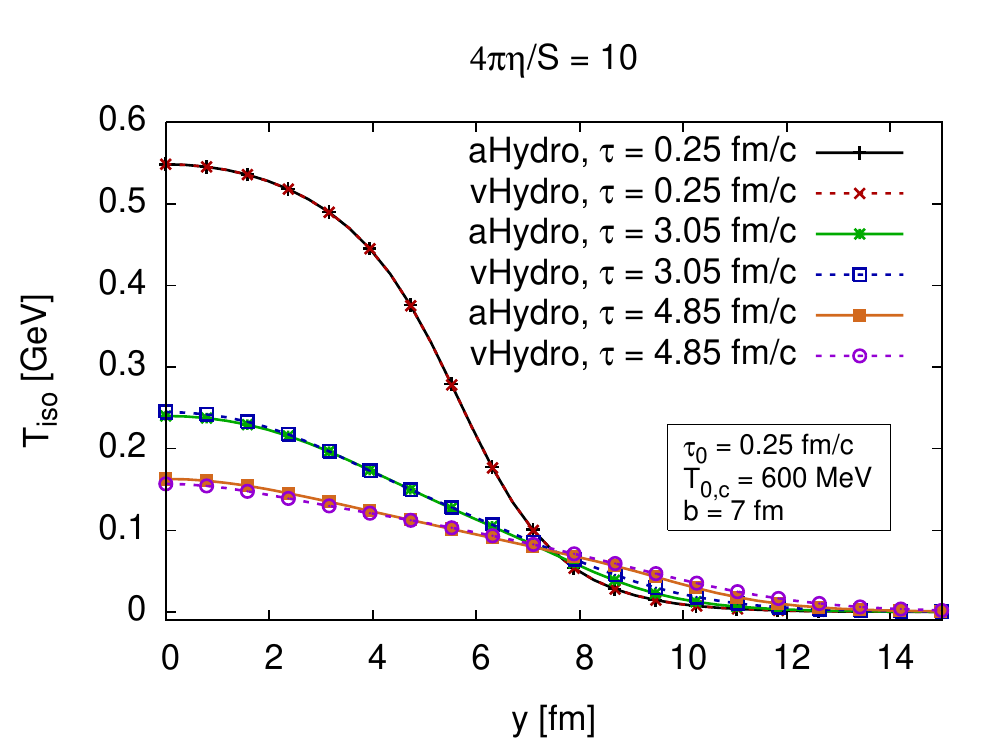}
\includegraphics[width=8.1cm]{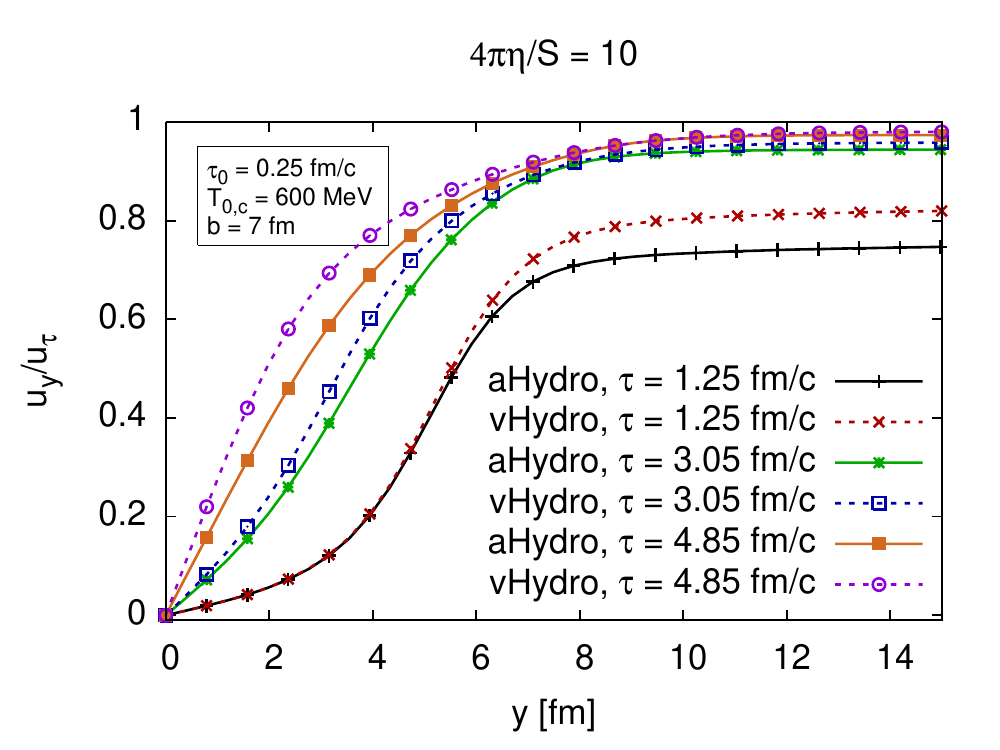}
\end{center}
\vspace{-8mm}
\caption{(Color online) Comparison of \ahydro isotropic temperature and flow profiles with 2nd-order viscous hydrodynamics code for
$4 \pi \eta/{\cal S}=10$ and $b=7$ fm.  Lattice size used was $109\times109$ with $a=0.394$ fm, $\epsilon=0.01$ fm/c, 
$\tau_0 = 0.25$ fm/c, $\Lambda_0 = $ 600 MeV, and $\xi_0 = 0$.  For the transverse profile Glauber binary collision scaling
was used.}
\label{fig:vCompare2}
\end{figure}

\begin{figure}
\begin{center}
\includegraphics[width=8.1cm]{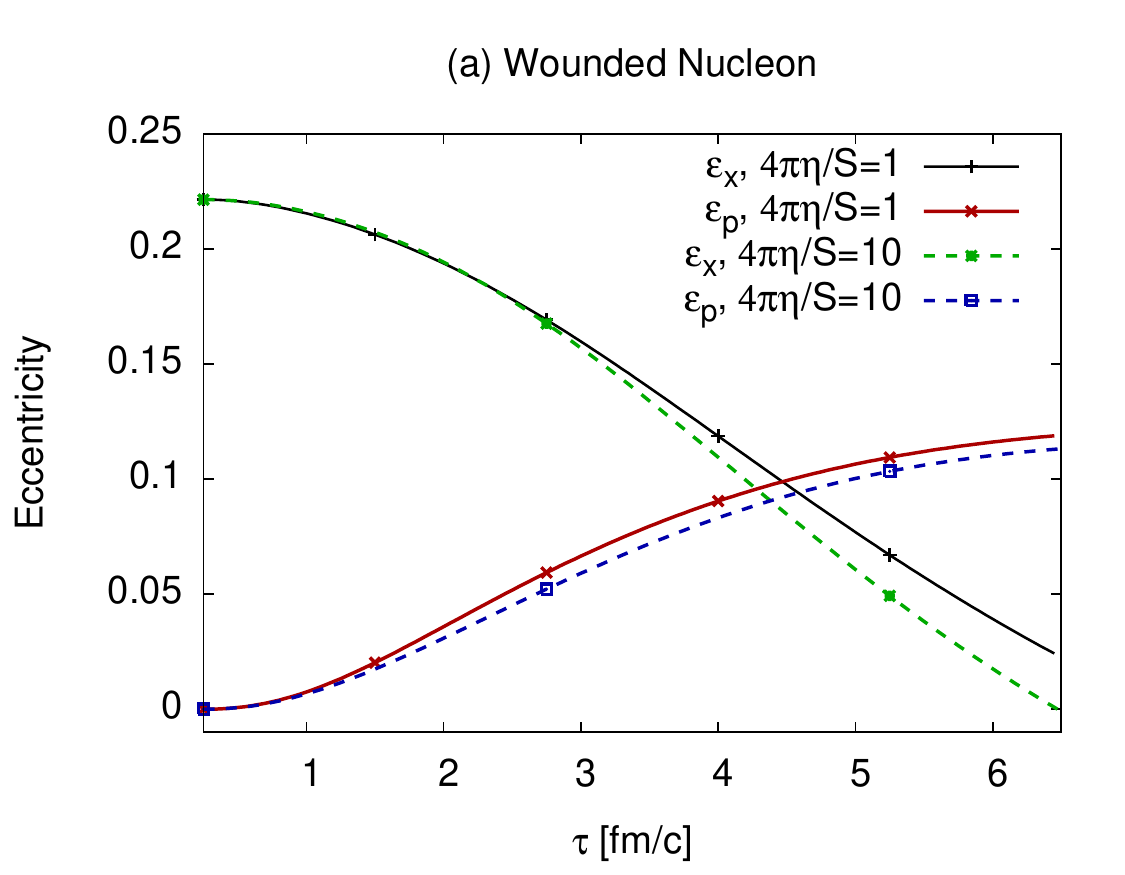}
\includegraphics[width=8.1cm]{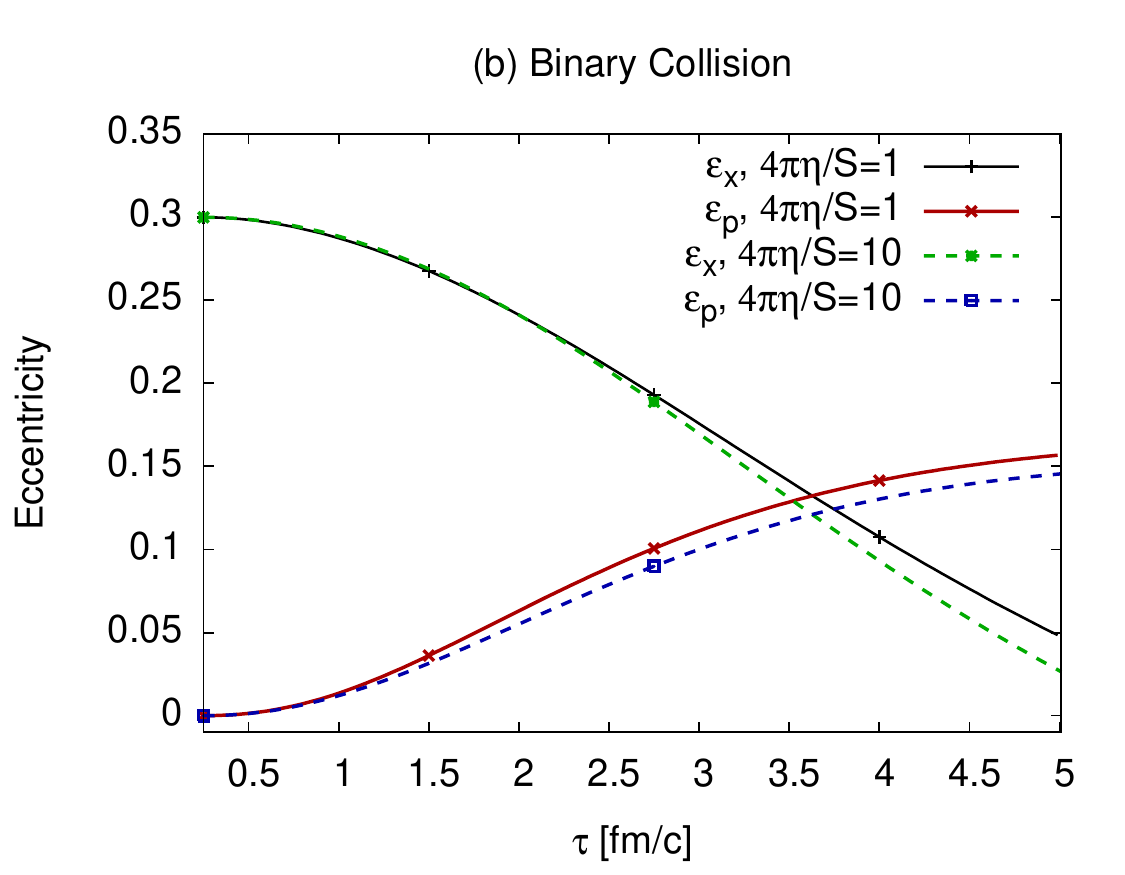}
\end{center}
\vspace{-8mm}
\caption{(Color online) Spatial and momentum eccentricities as a function of proper time for (a) a Glauber wounded-nucleon 
transverse profile and (b) a Glauber binary-collision transverse profile with $b=7$ fm, $\Lambda_0=T_0=0.6$ GeV, 
$\xi_0 = 0$, and $u_{\perp,0} = 0$ at $\tau_0 = $ 0.25 fm/c.  For the $4\pi\eta/{\cal S} = 1$ run 
we used $\Lambda_0=T_0=0.6$ GeV and for the $4\pi\eta/{\cal S} = 10$ run we used 
$\Lambda_0=T_0=0.576$ GeV for wounded-nucleon initial conditions and $\Lambda_0=T_0=0.584$ for binary-collision
initial conditions.  These adjustments were made in order to guarantee the same final particle number.  
In all cases we used the centered-differences algorithm with a lattice size of 
100 $\times$ 100, a lattice spacing of $a=0.4$ fm, and a RK4 temporal step size $\epsilon=0.01$ fm/c.}
\label{fig:eccCompare}
\end{figure}

\begin{figure}
\begin{center}
\includegraphics[width=5.65cm]{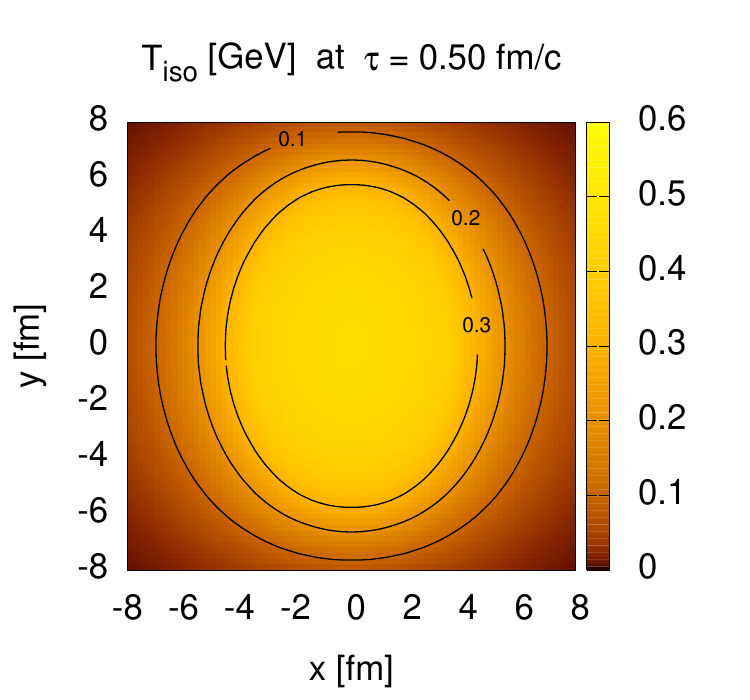}\hspace{-4mm}
\includegraphics[width=5.65cm]{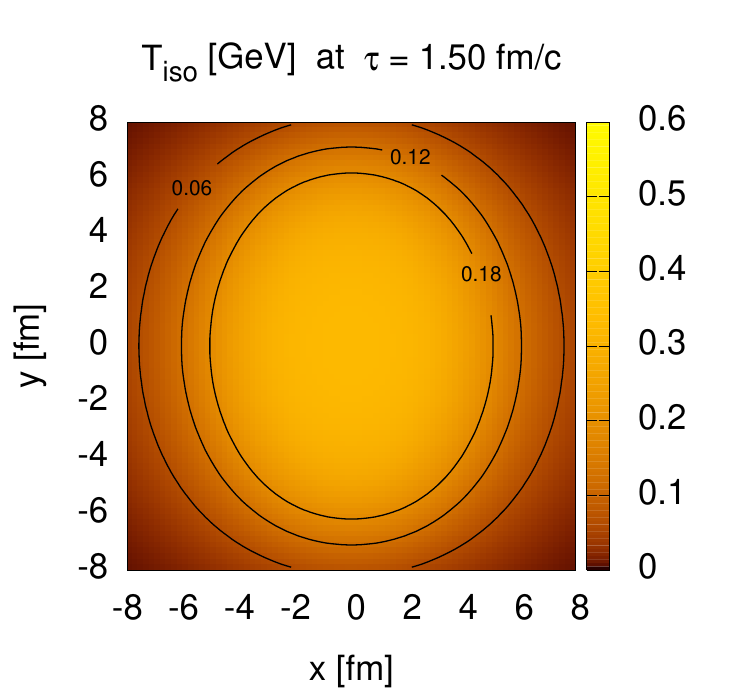}\hspace{-4mm}
\includegraphics[width=5.65cm]{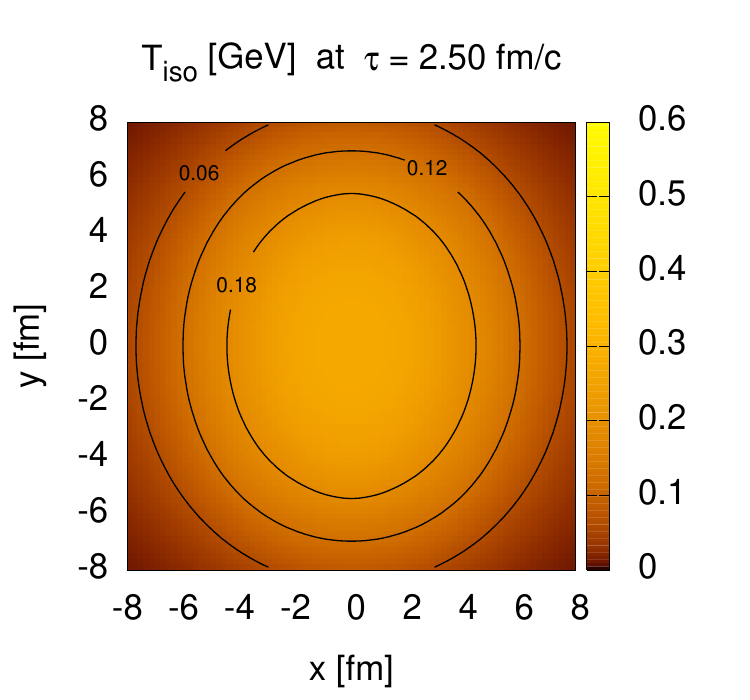}\\
\includegraphics[width=5.65cm]{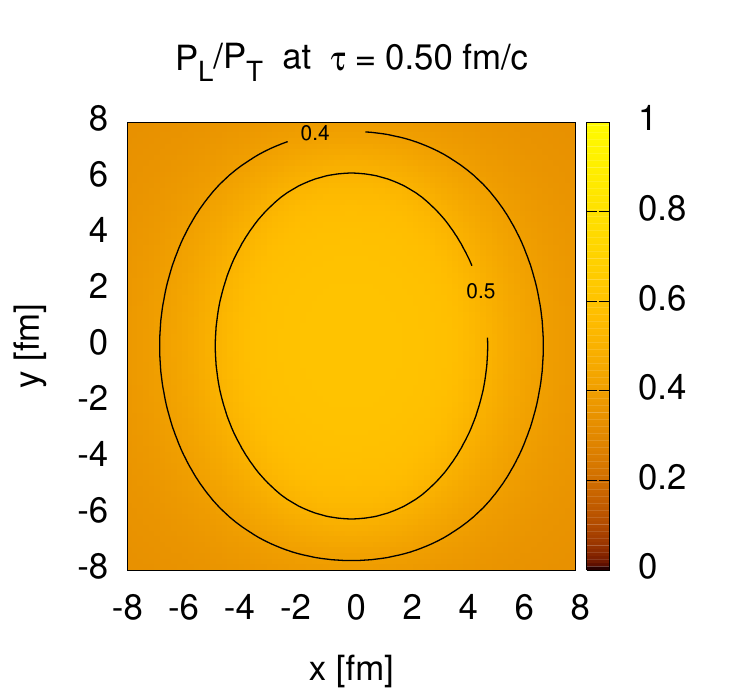}\hspace{-4mm}
\includegraphics[width=5.65cm]{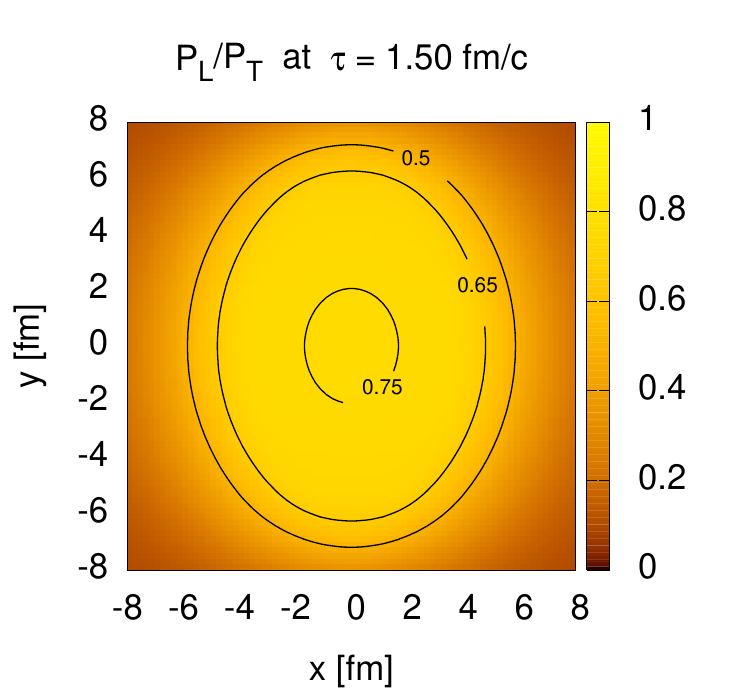}\hspace{-4mm}
\includegraphics[width=5.65cm]{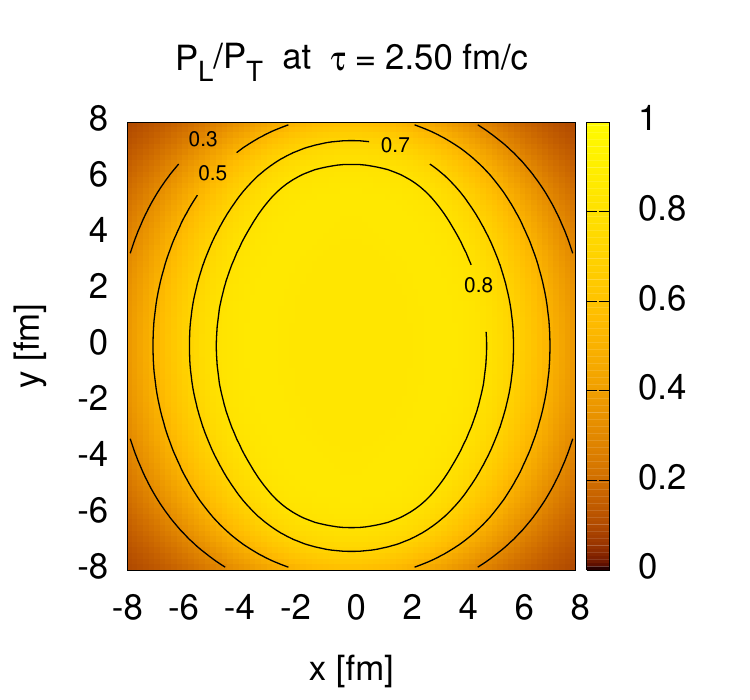}
\end{center}
\vspace{-8mm}
\caption{(Color online) Visualization of the isotropic temperature and pressure anisotropy at three different times after the
nuclear impact.  For these plots we assumed a non-central collision with $b=7$ fm, an isotropic Glauber wounded-nucleon 
profile, and a $b=0$ fm central temperature of 0.6 GeV at 0.25 fm/c.  For this plot we used a value of $4\pi\eta/{\cal S} = 1$ 
and a lattice size of $200\times200$ with a lattice spacing of $a=0.2$ fm and a RK4 temporal step size of $\epsilon=0.01$
fm/c.}
\label{fig:b7vis}
\end{figure}

\begin{figure}
\begin{center}
\includegraphics[width=5.65cm]{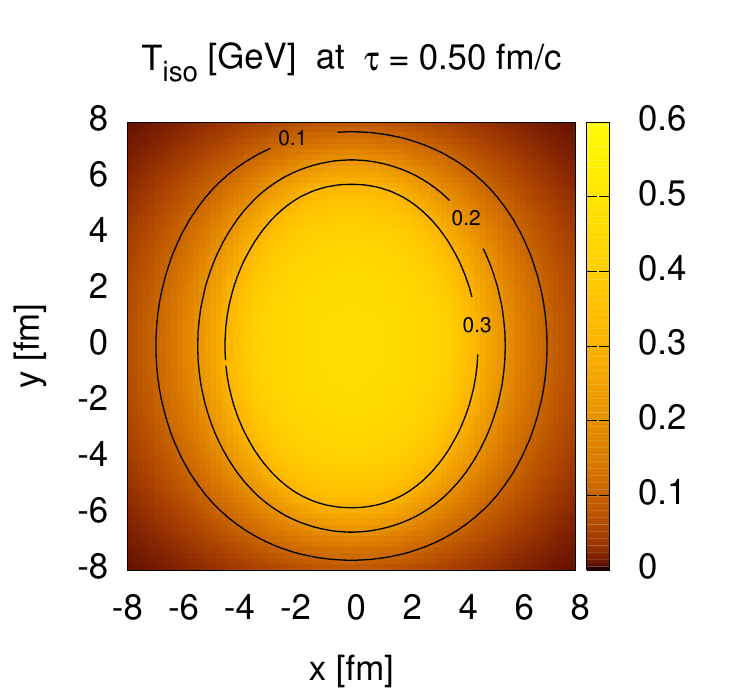}\hspace{-4mm}
\includegraphics[width=5.65cm]{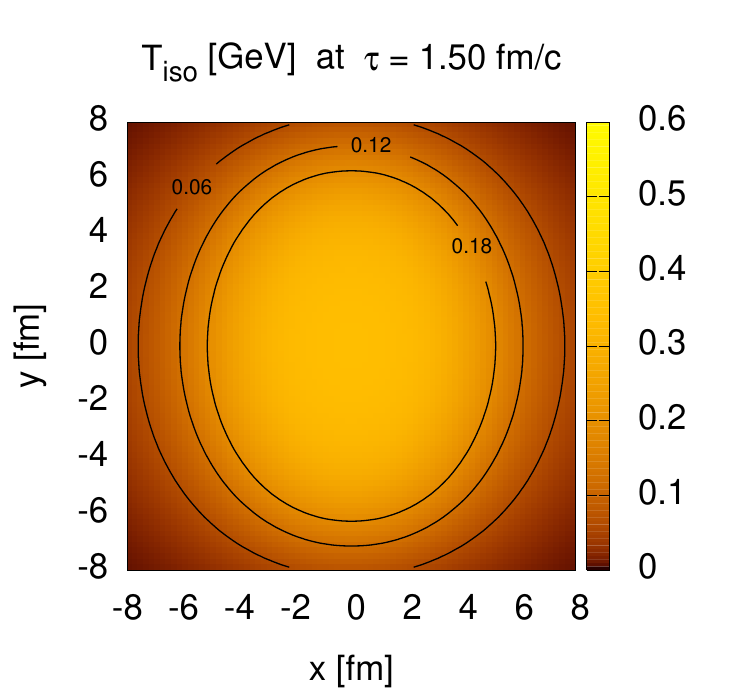}\hspace{-4mm}
\includegraphics[width=5.65cm]{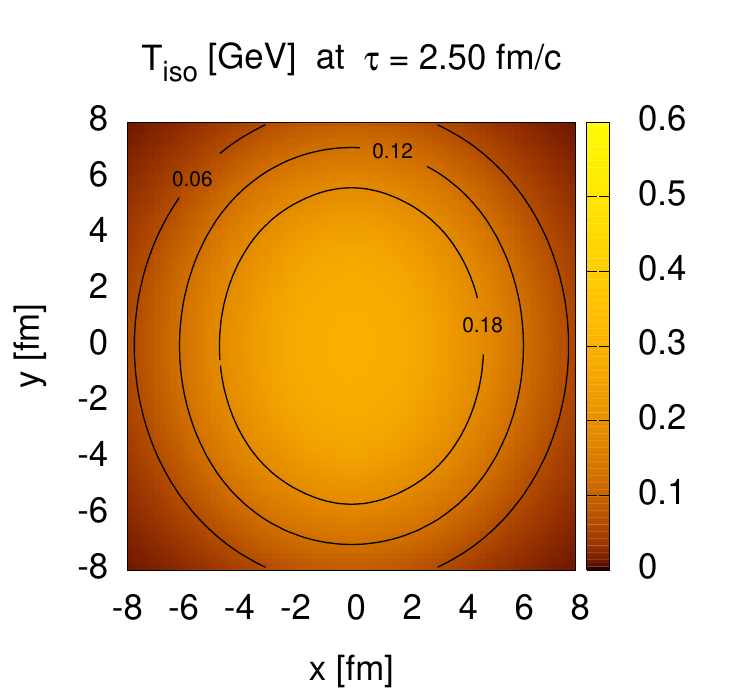}\\
\includegraphics[width=5.65cm]{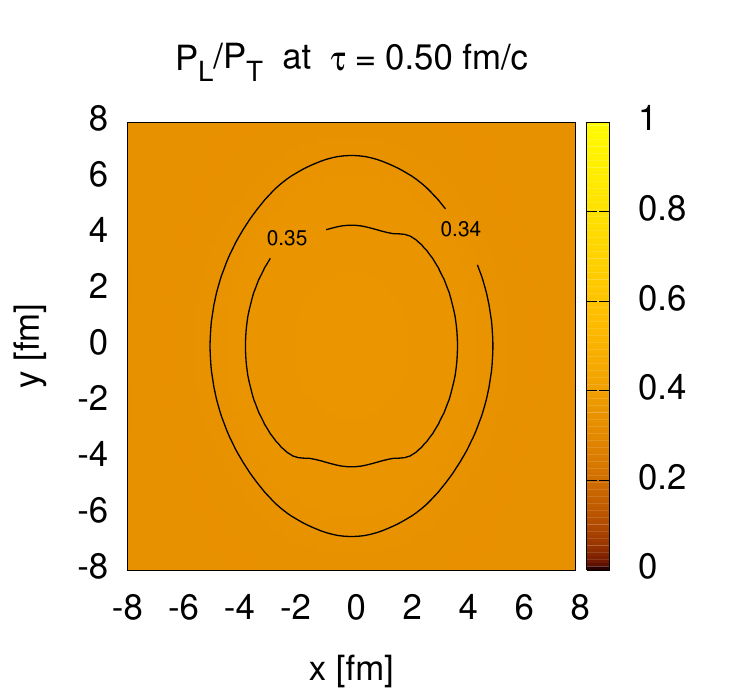}\hspace{-4mm}
\includegraphics[width=5.65cm]{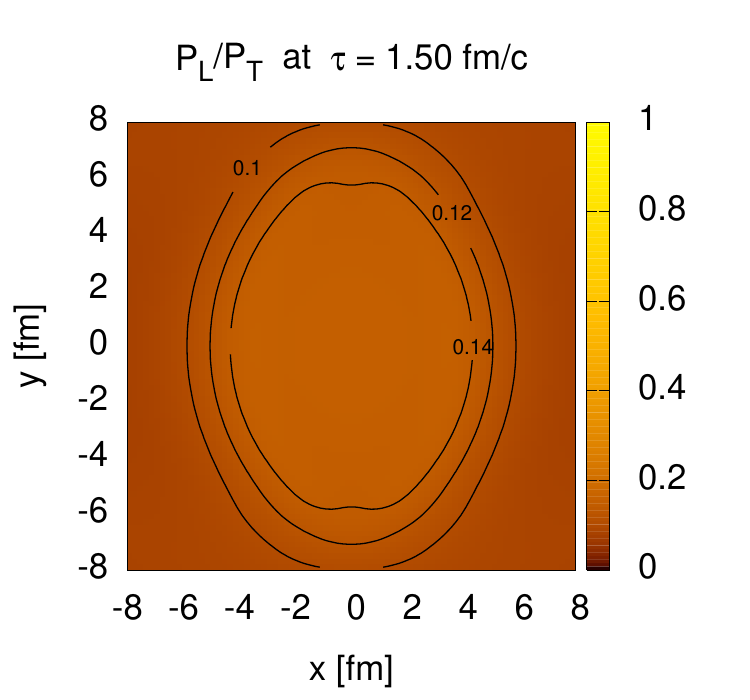}\hspace{-4mm}
\includegraphics[width=5.65cm]{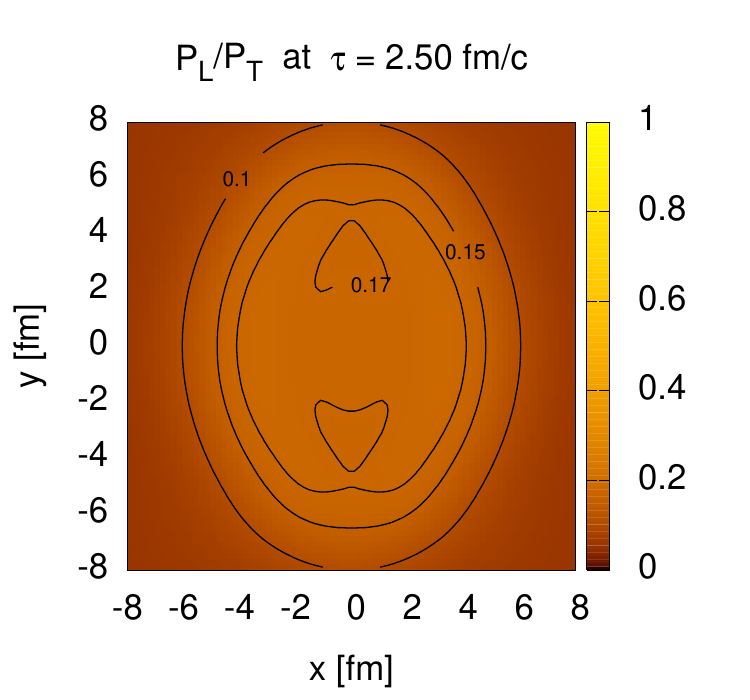}
\end{center}
\vspace{-8mm}
\caption{(Color online) Visualization of the isotropic temperature and pressure anisotropy at three different times after the
nuclear impact.  For these plots we assumed a non-central collision with $b=7$ fm, an isotropic Glauber wounded-nucleon profile, and a $b=0$ fm central temperature of 0.6 GeV at 0.25 fm/c.  For this plot we used a value of $4\pi\eta/{\cal S} = 10$ and a lattice size of $200\times200$ with a lattice spacing of $a=0.2$ fm and a RK4 temporal step size of $\epsilon=0.01$
fm/c.
}
\label{fig:b7lxivis}
\end{figure}

We concentrate on the ellipticities since, as we will see, large momentum-space anisotropies are developed during
the evolution of the system.  Such large momentum-space anisotropies cast doubt on the naive application of
Cooper-Frye \cite{Cooper:1974mv} and linearly-corrected Cooper-Frye \cite{Teaney:2003kp}.  We, therefore, 
postpone the implementation of freeze out until we can allow for large momentum-space anisotropies and, in the
meantime, focus on quantities that are independent of the freeze-out prescription.

\subsection{Smooth Initial Conditions}

We begin by presenting results using smooth initial conditions.  For numerical tests of the various algorithms
we refer the reader to App.~\ref{sec:numtests}.  Therein we show scalings with lattice spacing, box size, and comparisons of the different algorithms employed for both smooth and fluctuating initial conditions.

In order to demonstrate that \ahydro reproduces known 2nd-order viscous hydrodynamics results, in 
Figs.~\ref{fig:vCompare1} and \ref{fig:vCompare2} we compare the results of an \ahydro run with results 
obtained using the latest version of the code of Romatschke and Luzum \cite{Luzum:2008cw}.  In
Fig.~\ref{fig:vCompare1} we assumed $4 \pi \eta/{\cal S} = 0.1$ and in Fig.~\ref{fig:vCompare2} we assumed 
$4 \pi \eta/{\cal S} = 10$.  In both cases we show the isotropic temperature profile, $T_{\rm iso} = 
{\cal R}^{1/4}(\xi) {\cal E}_{\rm iso}(\Lambda)$, in the left panel and the ratio of the y-component of
the four velocity to the $\tau$-component in the right column.  As can be seen from Fig.~\ref{fig:vCompare1} 
there are only small differences at large radii in the case that the shear viscosity to entropy ratio is small.  
This
demonstrates that our code reproduces 2nd-order viscous hydrodynamics in the limit of small $\eta/{\cal S}$.
Fig.~\ref{fig:vCompare2} shows the case of large shear viscosity to entropy ratio.  In this case we see only small
deviations in the temperature profiles and substantial differences in the flow profiles.  We therefore expect the
\ahydro and 2nd-order viscous hydrodynamics frameworks to give different flow observables for large $\eta/{\cal S}$.
We note that corrections near the edges are expected even for small values of $\eta/{\cal S}$ and that the relative
magnitude of the \ahydro flow and the viscous hydrodynamics flow is to be expected:  since \ahydro generates 
larger longitudinal pressure than viscous hydrodynamics one expects diminished radial flow.  This pattern is 
also observed in simulations which use the lattice-boltzmann method \cite{Romatschke:2011hm}.

In Fig.~\ref{fig:eccCompare}a and \ref{fig:eccCompare}b we compare the spatial and transverse momentum-space 
eccentricities as a function of
proper time assuming two different values of the shear viscosity to entropy density ratio corresponding to typical
strong-coupling ($4\pi\eta/{\cal S} = 1$) and weak-coupling ($4\pi\eta/{\cal S} = 10$) values.  In Fig.~
\ref{fig:eccCompare}a we used smooth Glauber wounded-nucleon initial conditions and in Fig.~\ref{fig:eccCompare}b we used smooth
Glauber binary collision initial conditions.  In both figures we assumed $b=7$ fm, $\Lambda_0=T_0=0.6$ GeV, $\xi_0 = 
0$, and $u_{\perp,0} = 0$ at $\tau_0 = $ 0.25 fm/c and used the centered-differences algorithm with a lattice size of 
100 $\times$ 100, a lattice spacing of $a=0.4$ fm, and a temporal step size of 
$\epsilon=0.01$ fm/c.  In both cases RK4 with a temporal step size of $\epsilon=0.01$ fm/c was 
used for the updates.  As can be seen from these figures increasing the shear viscosity to entropy ratio by a factor of ten
only decreases the momentum-space eccentricity $\epsilon_p$ at 
5 fm/c by approximately 10\% in both cases shown.
We note, however, that the dynamical framework employed here, namely assuming that the local rest frame 
energy momentum tensor is azimuthally symmetric in momentum-space may underestimate the full effect of 
the shear viscosity.

In Figs.~\ref{fig:b7vis} and \ref{fig:b7lxivis} we present visualizations in the form of colormaps with contours of the
proper-time dependence of the isotropic temperature and the
pressure anisotropy defined by the ratio of the longitudinal and transverse pressures.   Fig.~\ref{fig:b7vis} shows
the case of $4 \pi \eta/S =1$ and Fig.~\ref{fig:b7lxivis} shows the case of $4 \pi \eta/S =10$.  In both cases we 
assumed a non-central collision
with $b=7$ fm, a Glauber wounded-nucleon profile, and a $b=0$ fm central temperature of $\Lambda_0 = T_0 = $
0.6 GeV at $\tau_0 = $ 0.25 fm/c.  A 
lattice size 
of $200\times200$ with a lattice spacing of $a=0.2$ fm and a RK4 temporal step size of $\epsilon=0.01$ fm/c was used 
in both cases.  As we can 
see from this figure the 
magnitude of the momentum-space anisotropies can be large in the center of the fireball and grows towards
the edges.  In Fig.~\ref{fig:b7vis} we see that assuming $4 \pi \eta/S =1$ at $\tau$ = 1.5 fm/c the center 
still has a 25\% momentum-space anisotropy and assuming $4 \pi \eta/S =10$ (Fig.~\ref{fig:b7lxivis}) one 
finds approximately 85\% momentum-space anisotropy at $\tau$ = 1.5 fm/c.  In fact, in the case of 
$4 \pi \eta/S =10$ the system is highly anisotropic during the entire evolution.  For such large shear viscosities
the {\sc aHydro} framework provides a dynamical framework which should be more reliable than the naive
application of 2nd-order viscous hydrodynamics.

\begin{figure}
\begin{center}
\includegraphics[width=10cm]{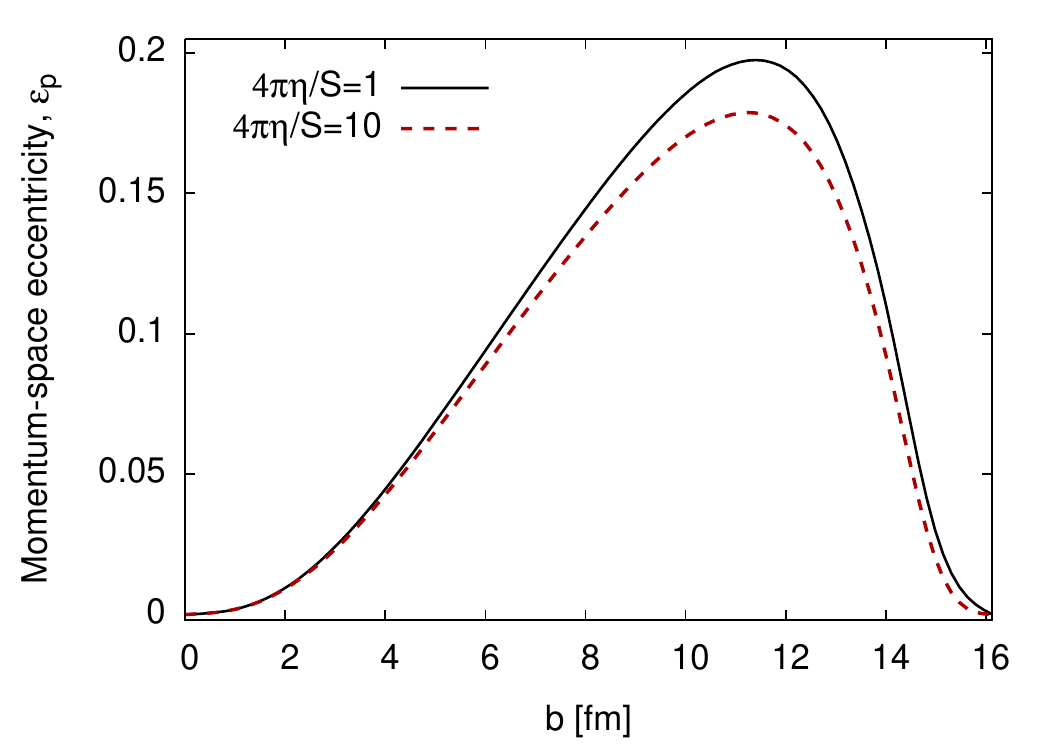}
\end{center}
\vspace{-8mm}
\caption{(Color online) Momentum eccentricity at the freeze-out time as a function of impact for an isotropic Glauber wounded-nucleon 
transverse profile with $\xi_0 = 0$, and $u_{\perp,0} = 0$ at $\tau_0 = $ 0.25 fm/c assuming 
$T_f = 0.15$ GeV.  For the $4\pi\eta/{\cal S} = 1$ run 
we used $\Lambda_0=T_0=0.6$ GeV as the central temperature and for the $4\pi\eta/{\cal S} = 10$ run we used 
$\Lambda_0=T_0=0.576$ GeV in order to guarantee the same final particle number.  
We used the centered-differences algorithm with a lattice size of 
200 $\times$ 200, a lattice spacing of $a=0.2$ fm, and a 
RK4 temporal step size $\epsilon=0.01$ fm/c.}
\label{fig:bdepLHC}
\end{figure}

In Fig.~\ref{fig:bdepLHC} we plot the momentum space eccentricity, $\epsilon_p$, at the ``freeze-out time'' $\tau_f$
as a function of the assumed impact parameter, $b$.  For this figure we used a Glauber wounded-nucleon 
transverse profile with $\xi_0 = 0$, and $u_{\perp,0} = 0$ at $\tau_0 = $ 0.25 fm/c assuming 
$4\pi\eta/{\cal S} = 1$ and $4\pi\eta/{\cal S} = 10$ and a freeze-out temperature of $T_f = 0.15$ GeV.  For the 
$4\pi\eta/{\cal S} = 1$ run we used $\Lambda_0=T_0=0.6$ GeV as the central temperature and for the 
$4\pi\eta/{\cal S} = 10$ run we used  $\Lambda_0=T_0=0.576$ GeV in order to guarantee the same final particle number.  
We used the centered-differences algorithm with a lattice size of 200 $\times$ 200, a lattice spacing of $a=0.2$ fm, 
and a RK4 temporal step size $\epsilon=0.01$ fm/c.  The freeze-out time $\tau_f$ was determined by finding
the time at which the maximum isotropic temperature $T_{\rm iso}$ 
dropped below the freeze-out temperature of $T_f = 0.15$ GeV.  This figure shows that
changing the assumed value of the shear viscosity to entropy ratio from one to ten only makes a difference of ~ 8\%
in the peak value of the momentum-space ellipticity.  We
should note, as a caveat which we will emphasize again in the conclusions, that because we assume that the energy
momentum tensor is azimuthally symmetric in the local rest frame this places us somewhere between a full
blown viscous hydrodynamical calculation and ideal hydrodynamics.  Therefore, firm conclusions will have to wait until results with a completely general ellipsoidal energy-momentum tensor are available.

\begin{figure}
\begin{center}
\includegraphics[width=5.6cm]{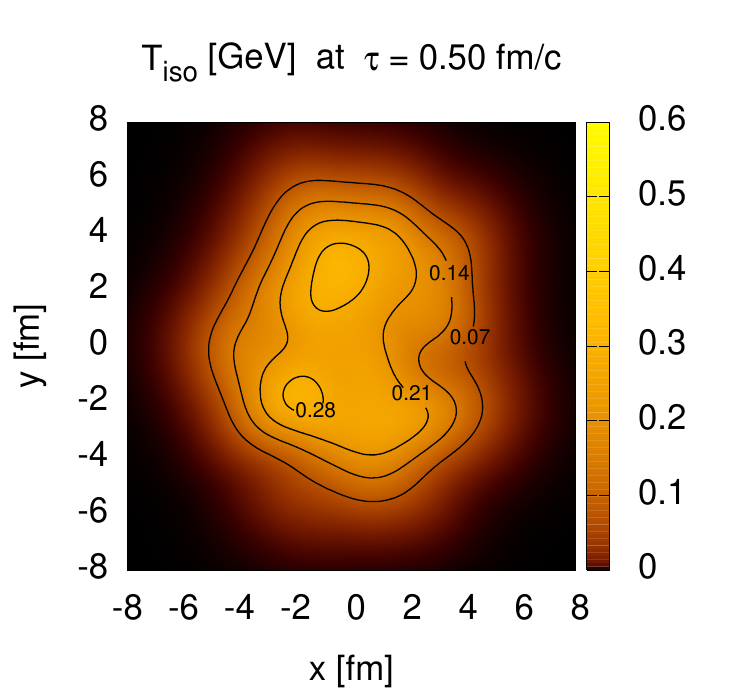}\hspace{-4mm}
\includegraphics[width=5.6cm]{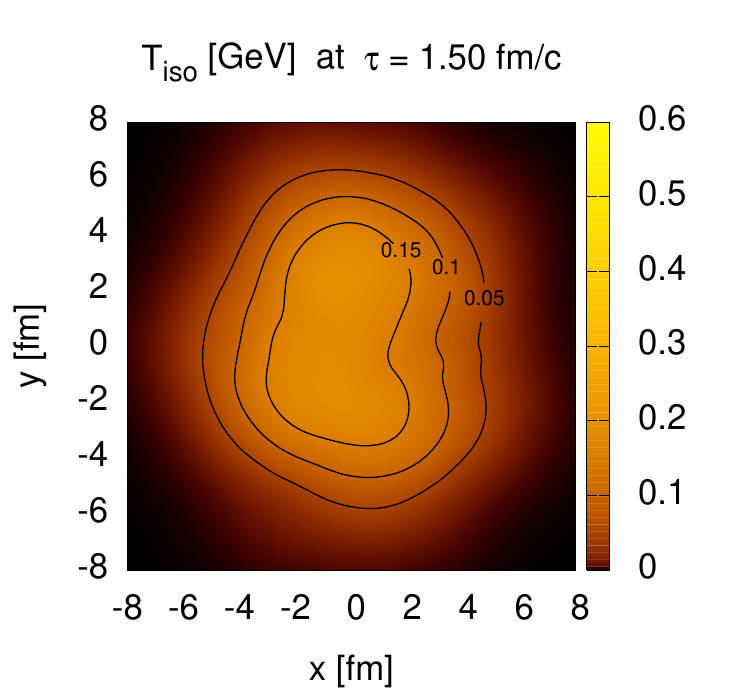}\hspace{-4mm}
\includegraphics[width=5.6cm]{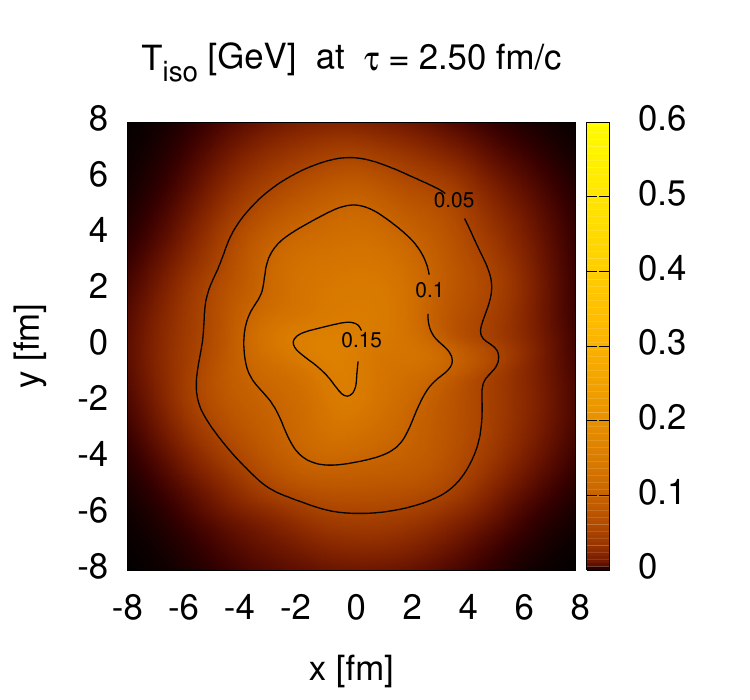}\\
\includegraphics[width=5.6cm]{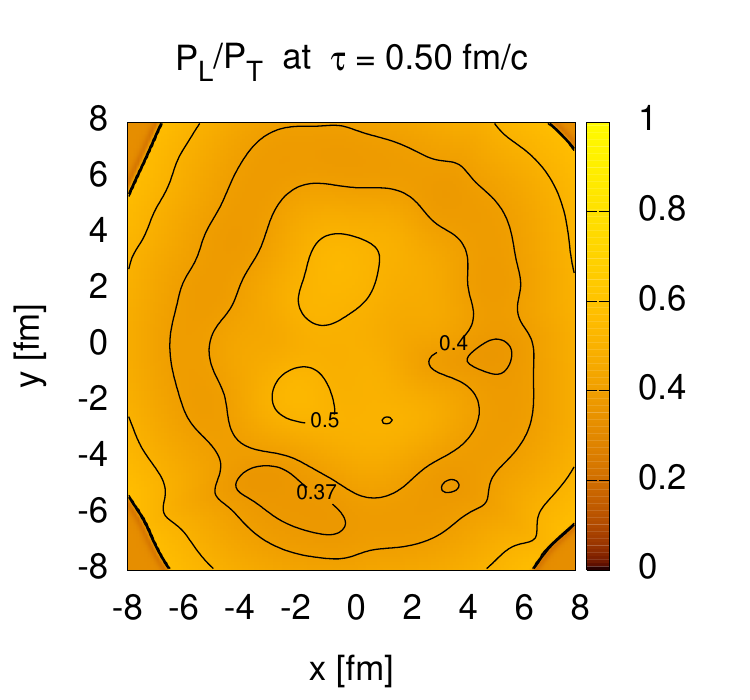}\hspace{-4mm}
\includegraphics[width=5.6cm]{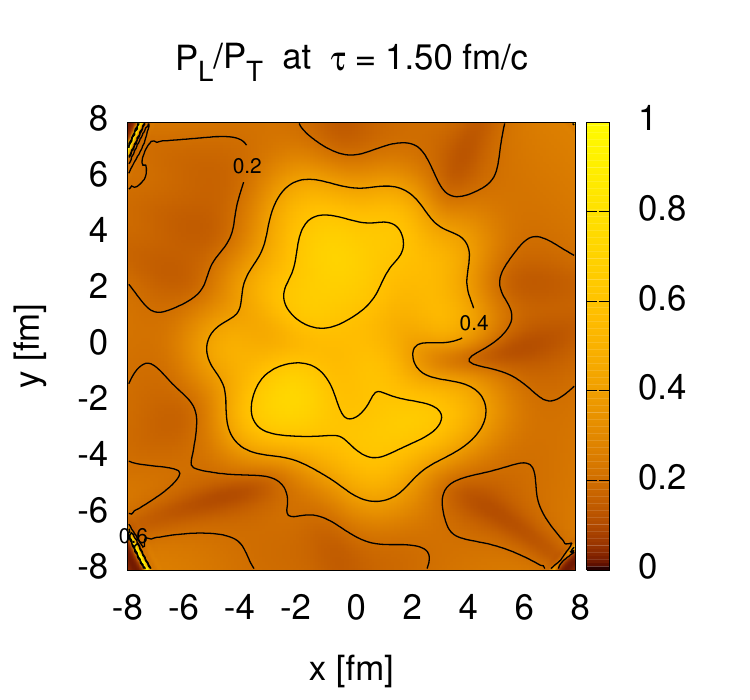}\hspace{-4mm}
\includegraphics[width=5.6cm]{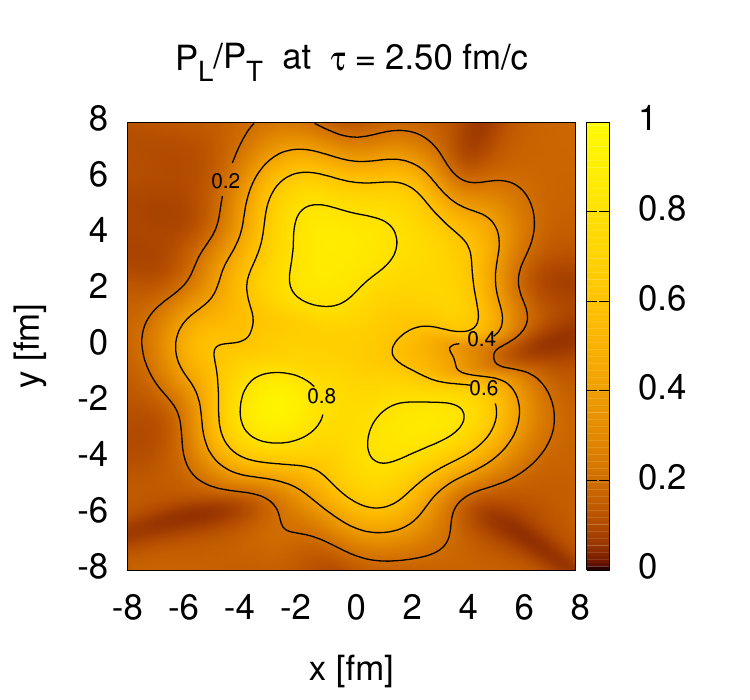}
\end{center}
\vspace{-8mm}
\caption{(Color online) Visualization of the isotropic temperature and pressure anisotropy at three different times after the
nuclear impact.  For these plots we assumed a collision centrality of $b=7$ fm with a sampled Monte-Carlo Glauber wounded-nucleon profile and an isotropic temperature of $T=0.6$ GeV at 0.25 fm/c. For this plot we used a value of 
$4\pi\eta/{\cal S} = 1$.  We used a lattice size of $200\times200$ with a lattice spacing of $a=0.2$ fm and a RK4 temporal step size of $\epsilon=0.01$ fm/c.
}
\label{fig:b7MCvis}
\end{figure}

\subsection{Fluctuating Initial Conditions}

For our fluctuating initial condition case we have implemented Monte-Carlo (MC) 
Glauber initial conditions \cite{Miller:2007ri}.
At a given impact parameter $b$ we statistically 
sample a Woods-Saxon distribution to determine the position of the nucleons in 
each colliding nuclei.  We then compute the transverse distance between each pair of nucleons from nuclei A and B 
and assume that they collide if the transverse distance between the centers of the nucleons being compared is
less than $d \equiv \sqrt{\sigma_{NN}/\pi}$.  If a collision is deemed to have occurred a two dimensional
gaussian with width $\sigma_0 = 0.46$ fm is added to the energy density.  We then adjust the overall scale to match
the smooth Glauber model results.

In Fig.~\ref{fig:b7MCvis} we present visualizations in the form of colormaps with contours of the
proper-time dependence of the isotropic temperature and the
pressure anisotropy defined by the ratio of the longitudinal and transverse pressures.   
In Fig.~\ref{fig:b7MCvis} we assumed a central collision $b=7$ fm with a sampled Monte-Carlo Glauber wounded-nucleon profile, an isotropic temperature of $\Lambda_0=T_0=0.6$ GeV at $\tau_0=$0.25 fm/c, and $4\pi\eta/{\cal S} = 1$.  We used a lattice size of $200\times200$ with a lattice spacing of $a=0.2$ fm and a RK4 temporal step size of $\epsilon=0.01$ fm/c.  As can be seen from this figure, fluctuations can induce large momentum-space anisotropies, particularly
in regions where the initial temperature is lower and therefore the relaxation rate is smaller.  In a 2nd-order 
viscous hydrodynamical approach one would have many ``spots'' with very large momentum-space anisotropies.
Note that Fig.~\ref{fig:b7MCvis} shows the case $4\pi\eta/{\cal S} = 1$ and we do not include a similar figure
for the case of $4\pi\eta/{\cal S} = 10$; however, we note that similarly to the case of smooth initial conditions,
for this large value of the shear viscosity to entropy ratio, one sees large persistent momentum-space 
anisotropies throughout the simulated region.

\section{Conclusions}
\label{sec:conclusions}

In this paper we studied the application of anisotropic hydrodynamics to the evolution of the matter created in 
relativistic heavy ion collisions.  We began by specifying a tensor basis for the energy-momentum tensor which was 
applicable when the system is azimuthally symmetric such that one has energy density, transverse pressure, and 
longitudinal pressure along the diagonal in the local rest frame.  Microscopically we were able to demonstrate that 
if one assumes local momentum-space azimuthal symmetry, it suffices to
introduce one scale $\Lambda$ and an anisotropy parameter, $\xi$, which controls the transverse-longitudinal
momentum-space anisotropy.  

We then used these results in the computation of moments
of the Boltzmann equation.  Using the zeroth and first moments of the Boltzmann equation we were able to determine 
dynamical equations for the plasma scale, $\Lambda$, anisotropy parameter, $\xi$, and the transverse flow components 
$u_x$ and $u_y$.  In order to solve the resulting partial differential equations we implemented three
differencing schemes:  centered differences, weighted LAX, and hybrid Kurganov-Tadmor.  The first method is
suitable for smooth initial conditions whereas the second two are required when one considers event-by-event
simulations.  Based on our analysis and benchmarks we find the weighted LAX scheme to be faster than the 
hybrid Kurganov-Tadmor scheme with both giving the same results within controllable numerical errors.

We showed through explicit solution of the resulting partial differential equations that the pressure
components remain positive definite and that plasma momentum-space anisotropies grow larger as one approaches the
transverse edge.  In addition, we studied fluctuating initial conditions and demonstrated that fluctuations can result in 
regions of high momentum-space anisotropy in the center of the simulated matter.  As a cross check we demonstrated
that in the limit of small $\eta/{\cal S}$ the solution of the \ahydro dynamical equations reproduces results 
from publicly
available 2nd-order viscous hydrodynamics codes.  For smooth initial conditions we demonstrated that, subject to the 
assumption of momentum-space azimuthal symmetry in the local
rest frame, one sees a relatively small variation of the final lab frame momentum-space eccentricity $\epsilon_p$ as
$\eta/{\cal S}$ is increased.  Drawing
quantitative conclusions from the results contained herein might be premature, however, since the impact of relaxing
the assumption of azimuthal isotropy of the energy momentum tensor in the local rest frame is unknown.  
Removing this assumption will result in what we will term ``ellipsoidal'' anisotropic hydrodynamics.  Work in this direction 
is currently underway.

We note in closing that there have been a number of authors studying the behavior of anisotropic plasmas
in strongly coupled gauge theories \cite{Janik:2008tc,Mateos:2011ix,Mateos:2011tv,Chernicoff:2012iq,Chernicoff:2012gu,%
Heller:2011ju,Heller:2012km,Giataganas:2012zy,Rebhan:2011ke,Rebhan:2011vd}.  The \ahydro framework 
agrees extremely well with existing 1st, 2nd, and 3rd order viscous hydrodynamical results 
which have been computed analytically
for strongly-coupled ${\cal N}=4$ supersymmetric Yang-Mills \cite{Booth:2009ct}.  It would be interesting to see
if any of the results contained herein could be used in the context of strongly-coupled theories in order to develop
useful phenomenological models.  One open question first
raised in Ref.~\cite{Rebhan:2011vd} concerns whether or not the breaking of rotational symmetry in momentum-space
requires the introduction of transverse and longitudinal transport coefficients.  Mathematically this would seem to be the
case in our formalism if one linearizes fluctuations around an anisotropic background.  Such possibilities will be explored in the future.
In the meantime, the progress made here opens up the possibility for phenomenological application to 
heavy ion observables such as collective flow, photon and dilepton production, quarkonium screening, jet 
energy loss, etc. in the presence of large momentum-space anisotropies.

\section*{Acknowledgements}
We thank Gabriel Denicol, Wojciech Florkowski, Sangjong Jeon, Harri Niemi, and Bj\"orn Schenke for 
useful conversations during
the preparation of this work.  M.~M. and M.~S. thank the H. Niewodnicza\'nski Institute of Nuclear Physics and 
the Frankfurt Institute of Advanced Studies where part of this work was done.   
M.S.  also thanks the Institute for Nuclear Theory at University of Washington for allowing him to 
participate in the INT program ``Gauge Field Dynamics In and Out of Equilibrium'' where the final
stages of this work were completed.
M.S. was supported by NSF grant 
No. PHY-1068765 and the Helmholtz International Center for FAIR LOEWE program.
M.~M. was supported by Ministerio de Ciencia e Innovacion of Spain under project FPA2009-06867-E.

\appendix

\section{Particle production in the (0+1)-dimensional case}
\label{sec:particleproduction}

In this appendix we discuss the issue of particle production in 2nd-order viscous hydrodynamics vs anisotropic
hydrodynamics.   To begin we note that there are two limits in which one expects particle production to go to
zero:  (a) the limit of ideal hydrodynamics and (b) the free-streaming limit.  For small but non-vanishing shear
viscosity we expect there to be additional particles associated with dissipation; however,
as the shear viscosity to entropy ratio increases we should see a maximum in the particle production since it
will eventually have to go to zero in the free-streaming limit.  In contrast, second-order viscous hydrodynamics 
predicts that the excess in particle production is a monotonically increasing function of the assumed value of 
$\eta/{\cal S}$.  

\begin{figure}[ht]
\begin{center}
\includegraphics[width=8cm]{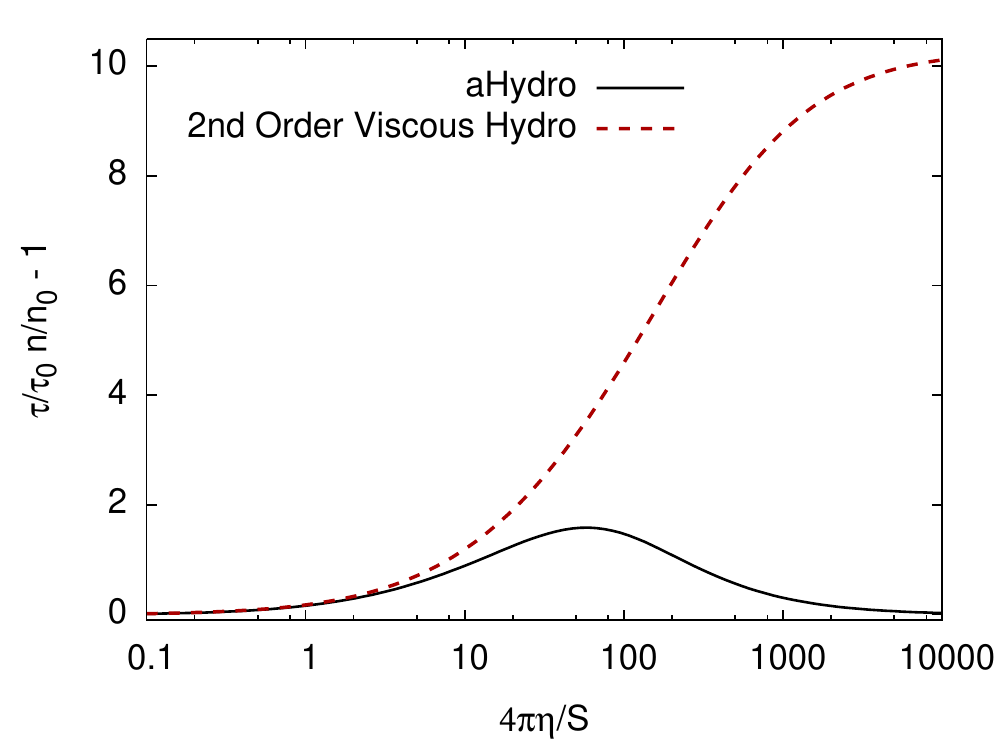}
\includegraphics[width=8cm]{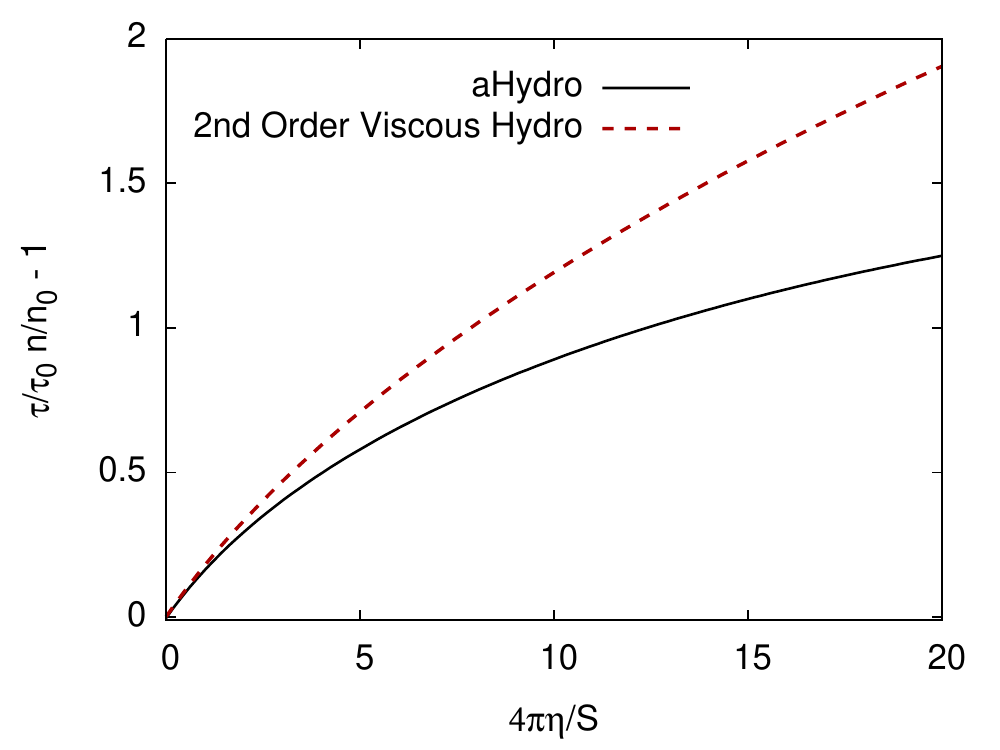}
\end{center}
\vspace{-8mm}
\caption{(Color online) Total particle number at $\tau = \tau_f$ as a function of the assumed value of the shear viscosity to 
entropy ratio.  For this figure we ignored transverse expansion making the system effectively (0+1)-dimensional
and we used initial values of $\Lambda_0=0.6$ GeV and $\xi_0 = 0$ at $\tau_0=0.25$ fm/c.}
\label{fig:entropyproduction}
\end{figure}

In order to demonstrate the difference quantitatively, in Fig.~\ref{fig:entropyproduction} we plot
the quantity $\tau/\tau_0 \, n/n_0 - 1$ at $\tau=\tau_f$ as a function of  $4\pi\eta/{\cal S}$.  
We used a freeze out temperature of $T_f = 150$ MeV to determine $\tau_f$.  This quantity should be zero if
there are no particles produced during the evolution. 
As can be seen
from these plots our expectations are confirmed, namely that one sees a maximum in entropy production at large
values of $4\pi\eta/{\cal S}$ with it returning to zero as $4\pi\eta/{\cal S}$ increases above this point.  Concentrating
on the zoomed plot in Fig.~\ref{fig:entropyproduction} one sees that for $4\pi\eta/{\cal S}=10$ 2nd-order viscous
hydrodynamics overestimates the entropy production by approximately 93\%.  We note that as the initial temperature
is lowered, the excess particle production obtained from 2nd-order viscous hydrodynamics becomes larger.  This 
will be important for phenomenology since one of the key constraints on $\eta/{\cal S}$ stems from having to 
reduce the assumed initial temperature in order to compensate for dissipative particle/entropy production.

\section{Numerical Tests}
\label{sec:numtests}

\begin{figure}
\begin{center}
\includegraphics[width=10cm]{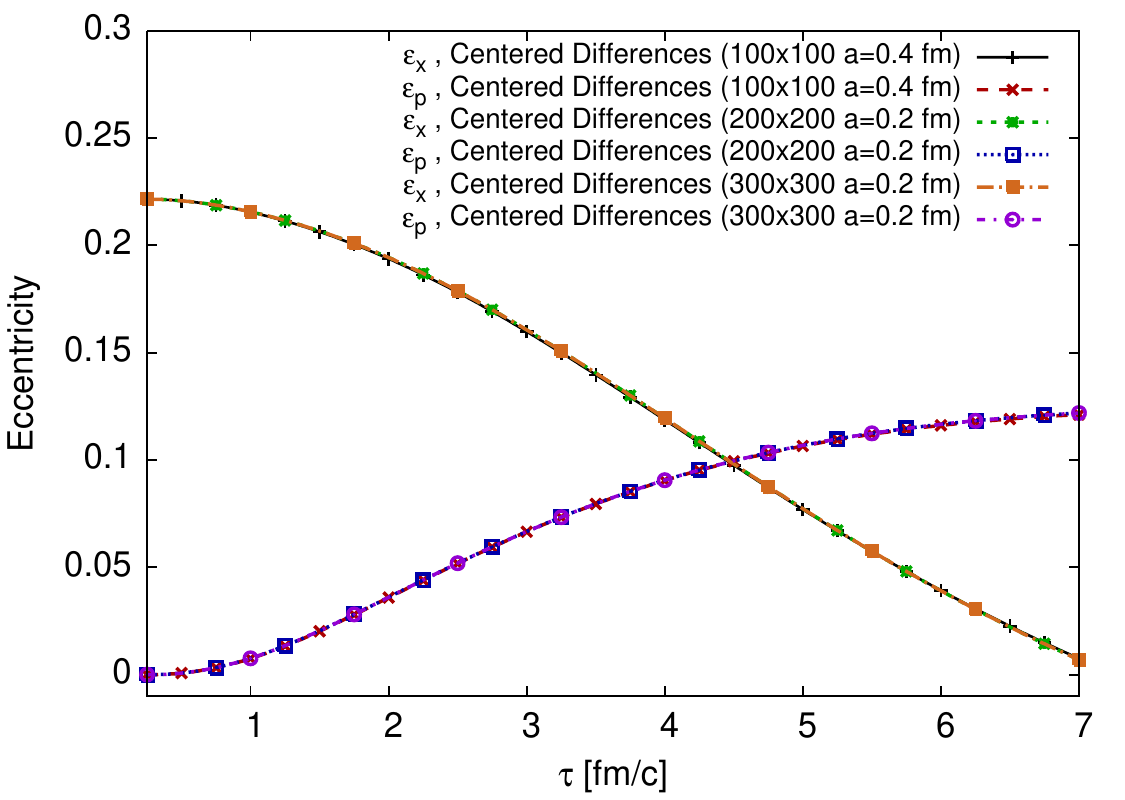}
\end{center}
\vspace{-8mm}
\caption{(Color online) Spatial and momentum eccentricities as a function of proper time for a smooth Glauber wounded-nucleon transverse
profile with $b=7$ fm, $\Lambda_0=T_0=0.6$ GeV, $\xi_0 = 0$, and $u_{\perp,0} = 0$ at $\tau_0 = $ 0.25 fm/c assuming 
$4\pi\eta/{\cal S} = 1$.  In all three cases we used a RK4 temporal step size of $\epsilon = 0.01$ fm/c.}
\label{fig:eccCDcompare}
\end{figure}

\begin{figure}
\begin{center}
\includegraphics[width=10cm]{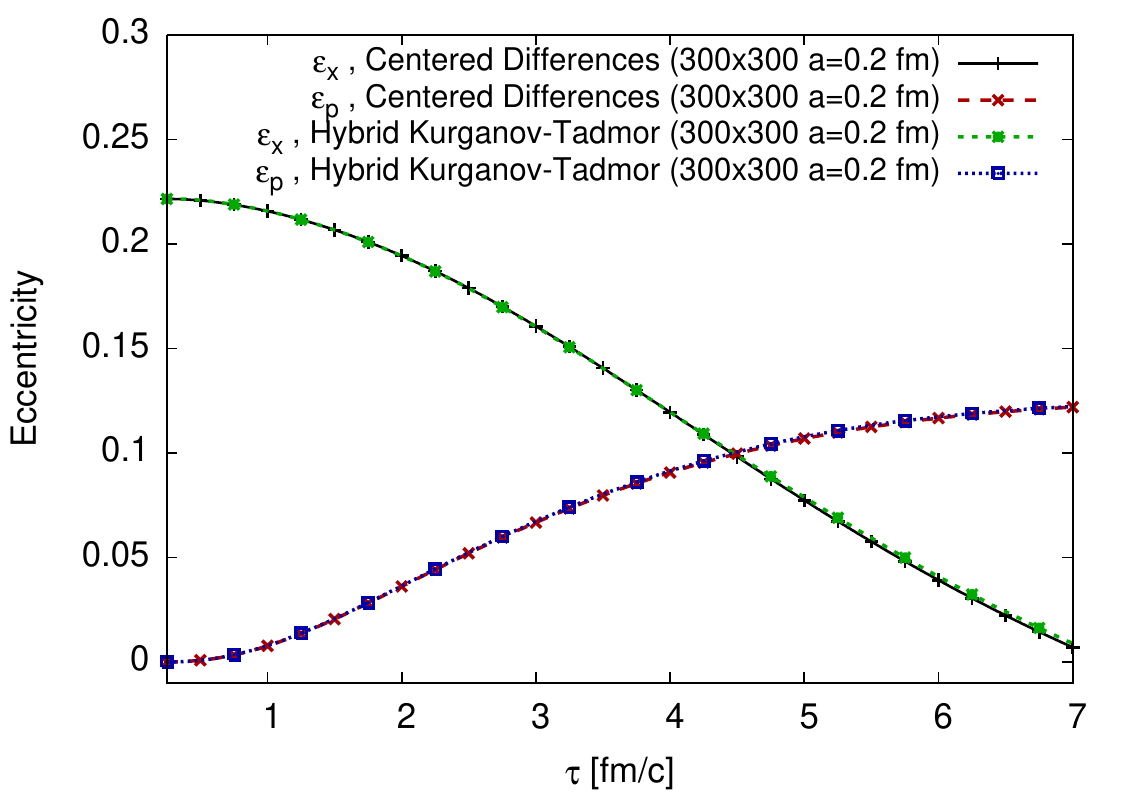}
\end{center}
\vspace{-8mm}
\caption{(Color online) Spatial and momentum eccentricities as a function of proper time for a smooth Glauber wounded-nucleon transverse
profile with $b=7$ fm, $\Lambda_0=T_0=0.6$ GeV, $\xi_0 = 0$, and $u_{\perp,0} = 0$ at $\tau_0 = $ 0.25 fm/c assuming 
$4\pi\eta/{\cal S} = 1$.  Here we compare the centered-differences and Hybrid Kurganov-Tadmor algorithms.}
\label{fig:eccCDHKTcompare}
\end{figure}

\begin{figure}
\begin{center}
\includegraphics[width=10cm]{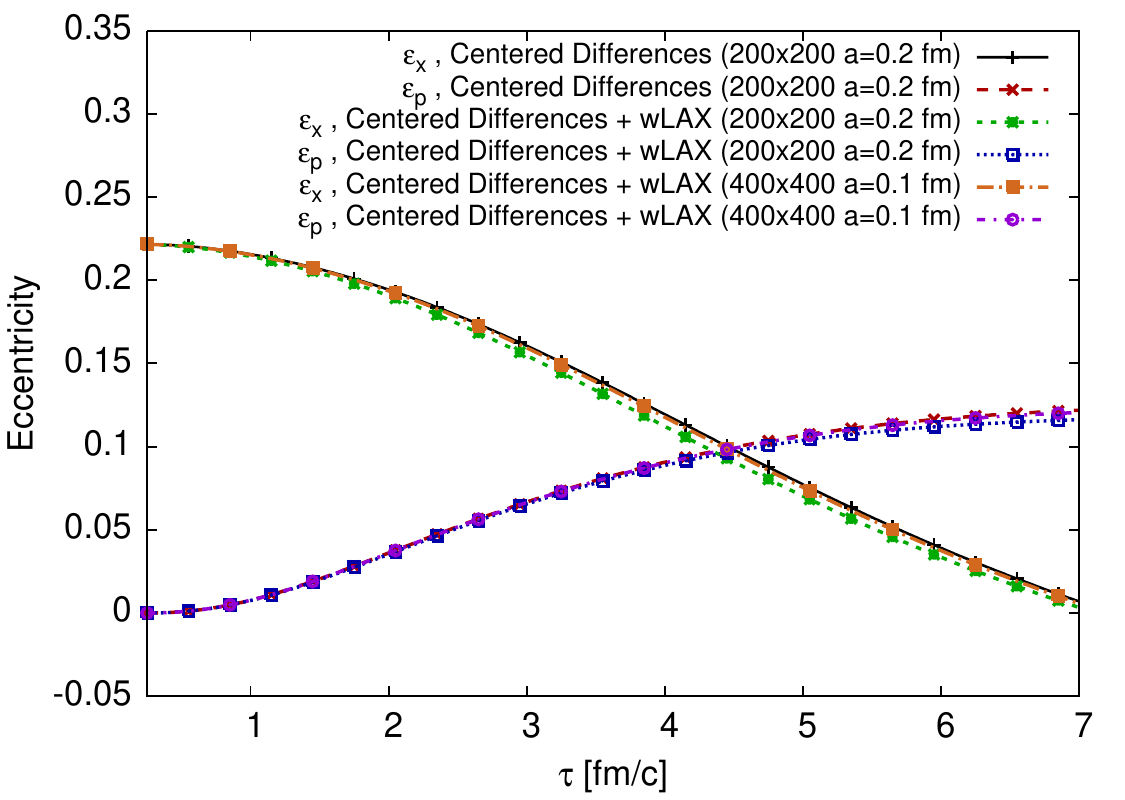}
\end{center}
\vspace{-8mm}
\caption{(Color online) Spatial and momentum eccentricities as a function of proper time for a smooth Glauber wounded-nucleon transverse profile with $b=7$ fm, $\Lambda_0=T_0=0.6$ GeV, $\xi_0 = 0$, and $u_{\perp,0} = 0$ at $\tau_0 = $ 0.25 fm/c assuming 
$4\pi\eta/{\cal S} = 1$.  Here we demonstrate the convergence of the wLAX algorithm with $\lambda=0.05$ to the result
obtained without any spatial smoothing as one 
decreases the lattice spacing.  In all cases RK4 with a temporal step size of $\epsilon=0.01$ fm/c was used.}
\label{fig:eccWLAXconverge}
\end{figure}

\begin{figure}
\begin{center}
\includegraphics[width=10cm]{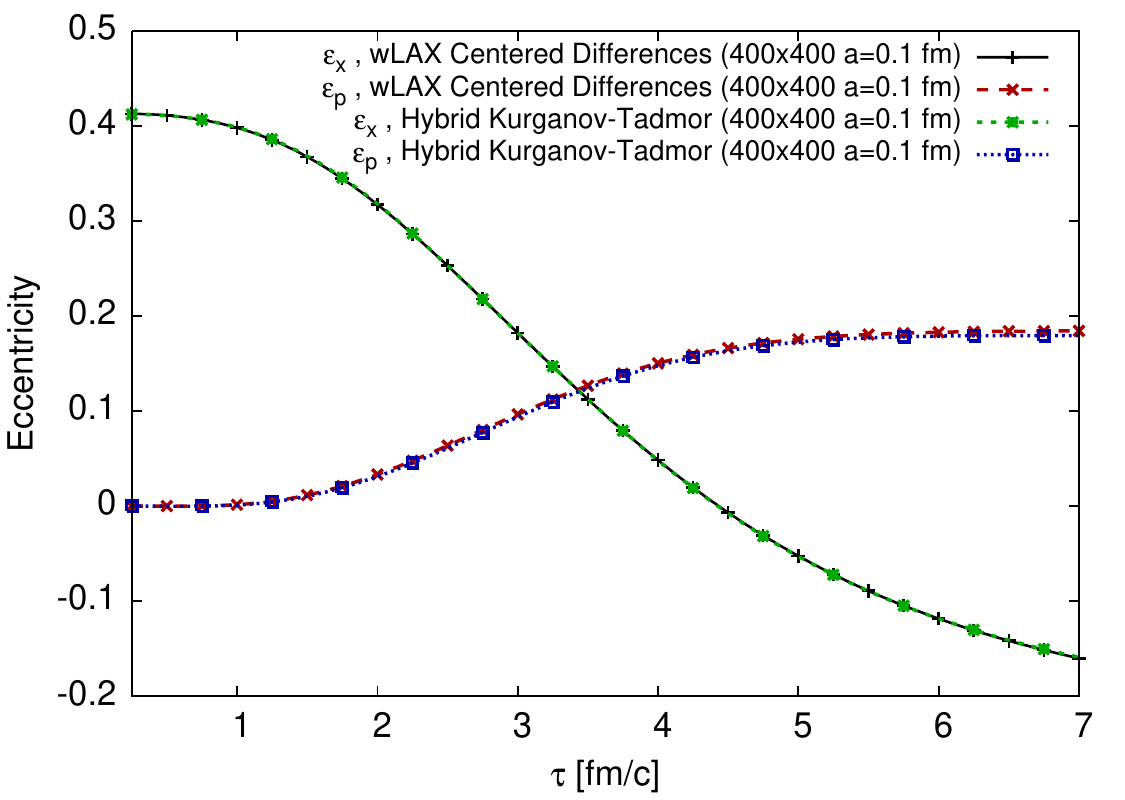}
\end{center}
\vspace{-8mm}
\caption{(Color online) Spatial and momentum eccentricities as a function of proper time for a sampled MC Glauber wounded-nucleon 
transverse profile with $b=7$ fm, $\xi_0 = 0$, and $u_{\perp,0} = 0$ at $\tau_0 = $ 0.25 fm/c assuming 
$4\pi\eta/{\cal S} = 1$.  Here we compare the hybrid Kurganov-Tadmor and wLAX algorithms.  
For the wLAX update we used RK4 with a temporal step size of $\epsilon=0.01$ fm/c.}
\label{fig:eccWLAXHKTfluccompare}
\end{figure}

In Fig.~\ref{fig:eccCDcompare} we show the time
evolution of the spatial and transverse momentum-space eccentricities as a function of proper time for a smooth Glauber 
wounded-nucleon 
transverse profile with $b=7$ fm, $\Lambda_0=T_0=0.6$ GeV, $\xi_0 = 0$, and $u_{\perp,0} = 0$ at $\tau_0 = $ 0.25 fm/c assuming $4\pi\eta/{\cal S} = 1$.  In all three cases we used a RK4 temporal step size of $\epsilon = 0.01$ fm/c. In 
this figure we have used the central-differences algorithm without wLAX smoothing
and compare the effect of 
varying the lattice spacing and lattice volume.  As can be seen from this figure, the systematics are well under
control in this case.  Knowing that the centered-differences algorithm systematics are under control we can now compare
with the hybrid Kurganov-Tadmor algorithm.  In Fig.~\ref{fig:eccCDHKTcompare} we show such a comparison for the
same conditions as shown in Fig.~\ref{fig:eccCDcompare}.  As can be seen from this figure the naive centered-differences
algorithm and the hybrid Kurganov-Tadmor algorithm give results that are indistinguishable by eye.

In Fig.~\ref{fig:eccWLAXconverge} we present the spatial and momentum eccentricities as a function of proper time 
for a smooth Glauber wounded-nucleon transverse profile with $b=7$ fm, $\Lambda_0=T_0=0.6$ GeV, $\xi_0 = 0$, 
and $u_{\perp,0} = 0$ at $\tau_0 = $ 0.25 fm/c assuming $4\pi\eta/{\cal S} = 1$.  In this plot we compare a
run with the unsmeared centered-differences algorithm and the wLAX algorithm with two different lattice spacings.
As can be seen from this figure the amount of numerical viscosity is small and can be reduced if one reduces the
lattice spacing.

To further illustrate the reliability of the wLAX algorithm in Fig.~\ref{fig:eccWLAXHKTfluccompare}
we compare a single MC Glauber wounded-nucleon run using both the wLAX and Hybrid Kurganov-Tadmor algorithms.
Both codes were initialized with the same sampled MC initial condition (a visualization of the evolution of this configuration
is shown in Fig.~\ref{fig:b7MCvis}).  As can be seen from this figure, wLAX and Hybrid Kurganov-Tadmor give virtually 
indistinguishable
results.  We point out in this context that the wLAX algorithm take much less time to complete a run giving it a
significant advantage when one wants to sample many different configurations.  Based on our benchmarks the
wLAX algorithm is approximately ten times faster than the Hybrid Kurganov-Tadmor algorithm.

\section{Boost invariant 1d dynamics - The Bjorken solution}
\label{sec:1dboost}

In this section we briefly review what happens when the system is boost invariant, 
homogeneous in the transverse directions, and has conserved particle number, i.e. $J_0 =0$.  
For this situation, it is convenient to switch to the comoving Milne coordinates defined as 
\begin{eqnarray}
t&=&\tau \cosh\varsigma \, , \nonumber \\
z&=&\tau \sinh\varsigma \, . 
\label{eq:com-coord}
\end{eqnarray}
In this coordinate system the metric $g_{\mu\nu}= {\rm diag}\,(1,-1,-1,-\tau^2)$. In addition, the local rest frame four-velocity simplifies to
\begin{equation}
u^\mu = (\cosh\varsigma,0,0,\sinh\varsigma) \, ,
\end{equation}
such that $u_\tau = 1$, $u_\varsigma = 0$, and we have
\begin{eqnarray}
D &=& u^\mu \partial_\mu = \partial_\tau \, , \nonumber \\
\theta &=& \partial_\mu u^\mu = \frac{1}{\tau} \, .
\label{eq:ddefs2}
\end{eqnarray}
By applying the last two expressions to the zeroth moment of the Boltzmann equation (\ref{eq:zerothmomgen}) for an isotropic plasma we obtain
\begin{equation}
\partial_\tau n = - \frac{n}{\tau} \, ,
\end{equation}
which has a solution of the form
\begin{equation}
n(\tau) = n_0\,\frac{\tau_0}{\tau} \, .
\end{equation}

If now we apply again the expressions given in Eq.~(\ref{eq:ddefs2}) to the first moment of the Boltzmann equation (Eq.\ref{eq:idealeq2}) one finds easily that
\begin{equation}
\partial_\tau{\cal E} + \frac{{\cal E} +{\cal P}}{\tau} = 0 \, .
\end{equation}
If the system has an ideal equation of state (EOS) then ${\cal E} = 3{\cal P}$ and one can further simplify this to
\begin{equation}
\partial_\tau{\cal E} = - \frac{4}{3} \frac{{\cal E}}{\tau} \, ,
\end{equation}
which has a solution
\begin{equation}
{\cal E}_{\rm ideal\;gas} = {\cal E}_0 \left(\frac{\tau_0}{\tau}\right)^{4/3} \, .
\end{equation}
If the system does not have an ideal EOS but instead has an equation of state corresponding to 
a constant speed of sound, i.e. $d{\cal P}/d{\cal E} = c_s^2$, then it follows that ${\cal P} = c_s^2 {\cal E}$
where we have fixed the constant by demanding that the pressure goes to zero when the energy density goes
to zero.  In this case one finds instead 
\begin{equation}
{\cal E} = {\cal E}_0 \left(\frac{\tau_0}{\tau}\right)^{1+c_s^2} \, ,
\end{equation}
which reduces to the ideal case when $c_s^2 = 1/3$.  If the EOS has  varying speed of sound then one
can express ${\cal P}$ in terms of an integral of the speed of sound.  Alternatively, one could calculate
the pressure and energy density separately for e.g. an ideal massive Boltzmann gas \cite{Romatschke:2011qp}
for which one finds
\begin{eqnarray}
{\cal E} &=& N_{\rm dof} \frac{e^{\mu/T} m^2 T}{2 \pi^2} 
\left[ 3 T K_2\left(\frac{m}{T}\right) + m K_1\left(\frac{m}{T}\right)\right] , 
\nonumber \\
{\cal P} &=& N_{\rm dof} \frac{e^{\mu/T} m^2 T^2}{2 \pi^2}K_2\left(\frac{m}{T}\right) ,
\nonumber \\
n &=&  \frac{\cal P}{T} \, , 
\end{eqnarray}
and
\begin{equation}
c_s^2(T,\mu=0) = \left(3 + \frac{m}{T} \frac{K_2(m/T)}{K_3(m/T)} \right)^{-1} \, .
\end{equation}
Note that the thermodynamic relations above are consistent with Bjorken scaling for the number
density, $n/n_0 = \tau_0/\tau$, for all values of $m$ in the case of isotropic hydrodynamics.  


\bibliography{ahydro}


\end{document}